\documentclass[preprint, prd,floatfix]{revtex4-2}
\usepackage{color}
\usepackage{hyperref}
\usepackage[english]{babel}
\usepackage{inputenc}
\usepackage{enumerate}
\usepackage{amsfonts}
\usepackage{amssymb}
\usepackage{amsmath}
\usepackage{dcolumn}
\usepackage{bm}
\usepackage{graphicx, graphics}
\usepackage{multirow}
\usepackage{graphicx}
\usepackage{natbib}
\usepackage{booktabs}
\usepackage{appendix}
\usepackage{makecell}
\begin{document}


\title{Holographic Calculation of Semileptonic Form Factors and the Isgur--Wise Function for \texorpdfstring{$B_{(s)}\to D_{(s)}$}{B(s)→D(s)} Decays}

\author{Miguel Angel Martin Contreras}
\email{miguelangel.martin@usc.edu.cn}
\affiliation{
 School of Nuclear Science  and Technology\\
 University of South China\\
 Hengyang, China\\
 No. 28, West Changsheng Road, Hengyang City, Hunan Province, China.
}

\author{Alfredo Vega}%
 \email{alfredo.vega@uv.cl}
\affiliation{%
 Instituto de F\'isica y Astronom\'ia, \\
 Universidad de Valpara\'iso,\\
 A. Gran Breta\~na 1111, Valpara\'iso, Chile
}

\begin{abstract}
We present a holographic analysis of the semileptonic decays $B \to D \ell \bar{\nu}_\ell$ and $B_s \to D_s \ell \bar{\nu}_\ell$ within the Softwall Model (SWM) and the Non-Quadratic Softwall Model (NQSWM). We extract the Isgur-Wise function from the overlap integral of bulk wavefunctions, with model parameters fixed by the heavy-light meson spectra. We impose Heavy Quark Symmetry (HQS) at zero recoil and address the apparent paradox of a non-conserved electroweak current dual to a massless bulk vector field. We argue that the non-conservation, arising from explicit flavor-- symmetry breaking by quark masses, does not generate an anomalous dimension. Hence the bulk field remains massless, and the non-conservation is \emph{encoded} in the independent boundary couplings $g_+$ and $g_-$. Our predictions for the branching fractions are in good agreement with LHCb data for $B_s \to D_s$ decays. The lepton-flavor universality ratios $R(D)$ and $R(D_s)$ are consistent with Standard Model and lattice QCD predictions. We find that the holographic models reproduce the universality of the Isgur-Wise function near zero recoil, while discrepancies at maximum recoil suggest that a more complete treatment of higher-order $1/m_Q$ corrections is needed.
 
\end{abstract}
\maketitle
\section{Introduction}

The semileptonic decays $B \to D \, \ell \, \bar{\nu}_\ell$ and $B_s \to D_s \, \ell \, \bar{\nu}_\ell$ provide an ideal laboratory for testing our understanding of strong interaction dynamics in the heavy-flavor sector. These transitions are particularly interesting because they allow for a direct extraction of the Cabibbo--Kobayashi--Maskawa (CKM) matrix element $|V_{cb}|$ and offer a stringent test of lepton flavor universality through the ratio $R(D^{(*)})$. Moreover, the theoretical description of these decays is greatly simplified by Heavy Quark Effective Theory (HQET), which organizes the non-perturbative QCD corrections in powers of $1/m_Q$ and predicts that all form factors reduce to a single universal function, the Isgur--Wise function $\xi(\omega)$ \cite{Isgur:1989vq, Isgur:1990kf}. Despite the remarkable successes of HQET, the functional form of $\xi(\omega)$ away from zero recoil remains a purely non-perturbative quantity that must be determined from first-principles methods such as lattice QCD \cite{Bernard:1993ct}, QCD sum rules \cite{Colangelo:1998fi, Lee:2002ka}, or phenomenological models \cite{Sadzikowski:1993iv}. In this work, we adopt the gauge/gravity duality, as realized in bottom-up AdS/QCD models, to compute the Isgur--Wise function (IWF) and the associated semileptonic form factors for the $B\to D$ and $B_s\to D_s$ transitions.

The holographic framework offers a powerful and intuitive approach to non-perturbative QCD, where hadrons are identified with normalizable modes of bulk fields propagating in a deformed anti-de Sitter (AdS) space. In particular, the Softwall Model (SWM) with a quadratic dilaton field successfully describes the light-flavor meson spectrum and electromagnetic form factors \cite{Karch:2006pv}. However, for heavy-light mesons, the mass hierarchy between the heavy quark and the light degrees of freedom motivates the use of a generalized dilaton profile, the Non-Quadratic Softwall Model (NQSWM) \cite{MartinContreras:2020cyg}, which introduces a deformation parameter $\alpha$ that controls the Regge trajectory curvature and accounts for the flavor-dependent dynamics of heavy quarkonium systems. By comparing the predictions of both models, we can assess the IWF's sensitivity to the confining-potential shape and the role of dilaton deformation in reproducing the heavy-quark limit.

We have addressed semileptonic form factor calculations in other holographic models and for different transitions. In Ref.~\cite{Ahmed:2023pod}, the authors investigate the $D^+ \to (\pi, K, \eta)\, \ell^+ \nu_\ell$, $D_s^+ \to (K, \eta)\, \ell^+ \nu_\ell$, and $D^+ \to K^* \ell^+ \nu_\ell$ and $D_s^+ \to K^* \ell^+ \nu_\ell$ transition form factors in the context of bottom-up 4-flavor holographic QCD. Although the authors obtain impressive results near the maximum recoil point, discrepancies emerge near zero recoil, possibly due to $1/m_c$ corrections, radiative QCD corrections (\emph{i.e.}, perturbative corrections in $\alpha_s$ to the weak current), or vector meson dominance. Ref.~\cite{Momeni:2022gqb} studies charm transitions in the bottom-up Hardwall model \cite{Polchinski:2002jw, Boschi-Filho:2002xih, Erlich:2005qh}. In the context of Light-Front Holographic QCD, Ref.~\cite{Ahmady:2013cga} determined the Light-Cone Distribution Amplitudes of the $\rho$ meson and used them to compute $B \to \rho$ transition form factors. These studies provide valuable benchmarks for our holographic calculations, as they cover a wide range of transitions and employ different holographic frameworks, allowing us to assess the model dependence of the form factor predictions.

In contrast to these studies, which focus primarily on the charm sector or employ specific holographic frameworks, our work provides a systematic comparison of the SWM and NQSWM for the $B \to D$ and $B_s \to D_s$ transitions within the HQS limit.

From the holographic point of view, form factors are defined from matrix amplitudes that have the structure $\langle H_2(p_2) | V^\mu | H_1(p_1) \rangle \sim f_+(q^2) (p^\mu + p'^\mu) + f_-(q^2) q^\mu$. Therefore, to extract the $f_\pm(q^2)$ form factors and the IWF, we impose Heavy Quark Symmetry (HQS) at zero recoil and fix the form-factor normalization according to HQET relations. This approach exploits the universality of the IWF and allows us to test the consistency of our holographic models with HQET. 

This manuscript is organized as follows: Section~\ref{sec:2} briefly defines the semileptonic form factors and the Isgur--Wise function. In Section~\ref{sec:3}, we present the holographic description of the $H_1 \to H_2$ transitions, where $H_X$ with $X=1,2$ are heavy-light mesons in a general bottom-up approach. We also compute the $f_\pm(q^2)$ form factors and the Isgur--Wise function $\xi(\omega)$ in the Softwall Model (SWM) and the Non-Quadratic Softwall Model (NQSWM). Section~\ref{sec.3a} first reviews the holographic dictionary for conserved currents, then discusses the apparent paradox emerging when we consider non-conserved weak currents with a massless bulk dual vector field. Section~\ref{sec:4} presents the parameter fixing and the specific holographic results. Finally, Section~\ref{sec:Concl} summarizes our main conclusions.

\section{Semileptonic decay and the Isgur--Wise function}\label{sec:2}
\begin{figure}[htbp]
    \centering
    \includegraphics[width=0.5\linewidth]{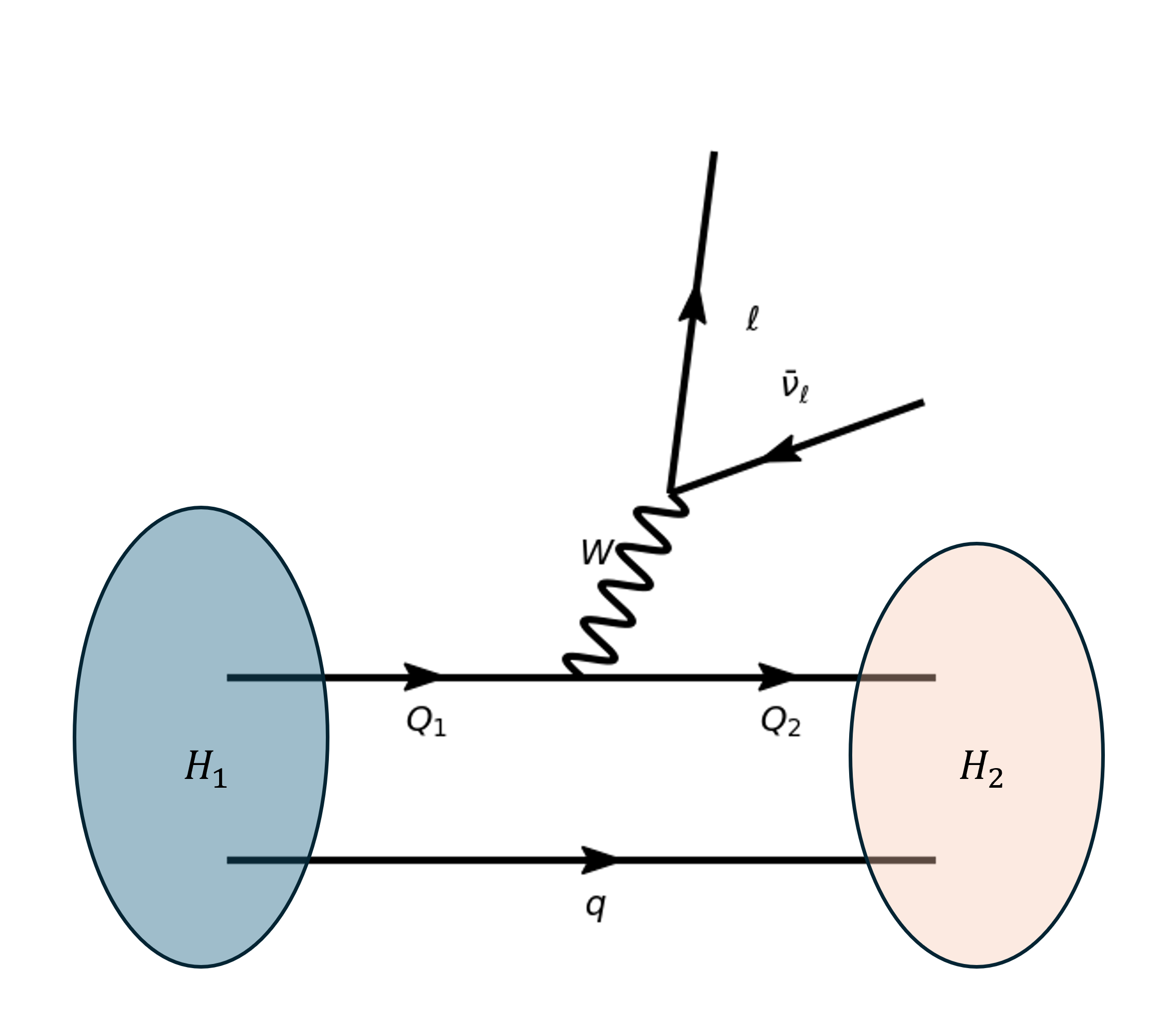}
    \caption{Feynman diagram showing the heavy-light meson decay $H_1\to H_2\,\ell\,\bar{\nu}_\ell$, displaying explicitly the weak current $\bar{Q}_2\,\gamma^\mu\,Q_1$. The spectator quark $q$ is either light $u$ and $d$ quark ($B\to D$) or a Strange quark $s$ ($B_s\to D_s$).}
    \label{fig:1}
\end{figure}

Semileptonic decays of heavy hadrons, such as $H_1 \to H_2 \, \ell \, \bar{\nu}_\ell$, provide a pristine laboratory for studying the strong-interaction dynamics of quark-flavor transitions. The weak charged current mediates the process, and its hadronic matrix elements encapsulate all non-perturbative QCD effects. In the limit where the masses of the initial- and final-state quarks are much larger than the QCD scale, the theoretical description simplifies considerably due to the emergence of heavy quark spin and flavor symmetries \cite{Wise:1997sg}. These symmetries dictate that all form factors for transitions within a given spin-symmetry multiplet are proportional to a single universal function of the velocity transfer, the Isgur--Wise function $\xi(\omega)$ \cite{Isgur:1989vq, Isgur:1990kf}.

For the transition between two pseudoscalar mesons, $H_1(p_1) \to H_2(p_2)$, the most general Lorentz-covariant decomposition of the vector current matrix element is

\begin{equation}
\langle H_2(p_2) | V^\mu | H_1(p_1) \rangle 
= f_+\left(q^2\right) \left(p_1^\mu + p_2^\mu - \frac{M_1^2 - M_2^2}{q^2} \,q^\mu\right)
+ f_0\left(q^2\right)\, \frac{M_1^2 - M_2^2}{q^2}\, q^\mu,
\label{eq:matrix_element_f0}
\end{equation}

\noindent where $q = p_1 - p_2$ is the momentum transfer, $M_1$ and $M_2$ are the masses of the initial and final hadrons, respectively, and $f_+(q^2)$ and $f_0(q^2)$ are the vector and scalar form factors. An equivalent parameterization is often employed,

\begin{equation}
\langle H(p_2) | V^\mu | H_1(p_1) \rangle  
= f_+\left(q^2\right) (p_1^\mu + p_2^\mu) + f_-\left(q^2\right) \,q^\mu,
\label{eq:matrix_element_fpm}
\end{equation}

\noindent with the two sets of form factors related by

\begin{equation}\label{eq:f0_relation}
f_0\left(q^2\right) = f_+\left(q^2\right) + \frac{q^2}{M_1^2 - M_2^2} f_-\left(q^2\right).
\end{equation}

The momentum transfer varies within the physical range $m_\ell^2 \leq q^2 \leq (M_1 - M_2)^2 \equiv q^2_{\text{max}}$.

In the heavy-quark limit ($m_{Q_1}, m_{Q_2} \to \infty$), the form factors reduce to a single universal function of the velocity transfer $\omega = v \cdot v'$, where $v$ and $v'$ are the four-velocities of the initial and final hadrons \cite{Isgur:1989vq, Isgur:1990kf}:

\begin{equation}
f_\pm(q^2) = \pm\frac{M_1 \pm M_2}{2\sqrt{M_1 M_2}} \, \xi(\omega),
\label{eq:iw_form_factors}
\end{equation}

\noindent where $\xi\left(\omega\right)$ is the so-called \emph{Isgur--Wise function}, and $\omega$ is defined as

\begin{equation}
\omega = \frac{M_1^2 + M_2^2 - q^2}{2 M_1 M_2}.
\label{eq:omega_def}
\end{equation}

A direct consequence of heavy quark symmetry at the zero-recoil point ($\omega = 1$, $q^2 = q^2_{\text{max}}$), where the final-state meson is at rest relative to the initial one, is the normalization of the Isgur--Wise function \cite{Isgur:1989vq, Isgur:1990kf}:

\begin{equation}
\xi(1) = 1.
\label{eq:iw_normalization}
\end{equation}

This leads to the non-trivial normalization of the form factors at maximum $q^2$ \cite{KimPham, Narison}:

\begin{align}
f_\pm(q^2_{\text{max}}) &= \pm\frac{M_1 \pm M_2}{2\sqrt{M_1 M_2}}, \label{eq:fpm_max} \\
f_0(q^2_{\text{max}}) &= \frac{2\,\sqrt{M_1\,M_2}}{M_1 + M_2}. \label{eq:f0_max}
\end{align}

These expressions will help constrain the corresponding holographic form factors. 

Let us now consider the differential decay rate for the process $H_1(p_1) \to H_2(p_2) \, \ell(k_1) \, \bar{\nu}_\ell(k_2)$, accounting for a massive lepton, which is given by \cite{Burdman:1993es}

\begin{multline}\label{eq:diff_decay_rate}
    \frac{d\Gamma}{dq^2} = \frac{G_F^2 \,\left|V_{\text{CKM}}\right|^2}{192\, \pi^3\,M_1^3}
\left(1-\frac{m_\ell^2}{q^2}\right)
\left[
\left(1+\frac{m_\ell^2}{2\,q^2}\right) \lambda^{3/2}\left(q^2\right) \,\left|f_+\left(q^2\right)\right|^2\right.\\
+ \left.\frac{3\,m_\ell^2}{2\, q^2} \left(M_1^2 - M_2^2\right)^2 \lambda^{1/2}\left(q^2\right) \left|f_0\left(q^2\right)\right|^2
\right],
\end{multline}

\noindent where $V_{\text{CKM}}$ is the relevant \emph{Cabibbo--Kobayashi--Maskawa matrix element}, $G_F$ is the Fermi constant, and

\begin{equation}
\lambda(q^2) = (M_1^2 + M_2^2 - q^2)^2 - 4 M_1^2 M_2^2
\label{eq:lambda}
\end{equation}

\noindent is the \emph{Källén function}. For light leptons ($\ell = e, \mu$), the lepton mass can be neglected, simplifying the expression to

\begin{equation}
\frac{d\Gamma}{dq^2} = \frac{G_F^2\, \left|V_{\text{CKM}}\right|^2}{192\,\pi^3\,M_1^3} \, \lambda^{3/2}\left(q^2\right) \left|f_+\left(q^2\right)\right|^2.
\label{eq:diff_decay_rate_light}
\end{equation}

The branching fraction is then obtained by integrating the differential rate over the allowed $q^2$ range and multiplying by the lifetime $\tau(H_1)$ of the initial-state meson:

\begin{equation}
\mathcal{B}(H_1 \to H_2 \, \ell \, \bar{\nu}_\ell)
= \tau(H_1) \int_{m_\ell^2}^{q^2_{\text{max}}} dq^2 \, \frac{d\Gamma}{dq^2}.
\label{eq:branching_ratio}
\end{equation}

A particularly interesting observable is the ratio of branching fractions for tauonic and light-lepton semileptonic decays,

\begin{equation}\label{eq:R_definition}
R(H_1\to H_2) \equiv 
\frac{\mathcal{B}(H_1 \to H_2 \, \tau \, \bar{\nu}_\tau)}
{\mathcal{B}(H_1 \to H_2 \, \ell \, \bar{\nu}_\ell)},
\qquad \ell = e, \mu,
\end{equation}

which provides a stringent test of lepton flavor universality (LFU) and a sensitive probe of potential new physics beyond the Standard Model (SM). In the SM, the couplings of the charged weak current to all lepton flavors are identical, a feature known as lepton flavor universality. The main advantage of the ratio $R(H_1\to H_2)$ is that the dominant hadronic uncertainties, encoded in the form factors, largely cancel, making it an exceptionally clean observable for probing physics beyond the SM.

The Belle, BaBar, and LHCb collaborations have extensively studied deviations from the SM predictions for these ratios, particularly in the channels $B \to D$ and $B \to D^*$ \cite{HFLAV:2024ctg}. For the pseudoscalar channel $B \to D$, the current world average reported by the Heavy Flavor Averaging Group (HFLAV) is $R(D) = 0.342 \pm 0.026$ \cite{HFLAV:2024ctg}, while the most precise SM predictions, obtained from lattice QCD, give $R(D)_{\rm SM} = 0.298 \pm 0.004$ \cite{Na:2015kha, FermilabLatt:2015ilb}. The discrepancy between the experimental average and the SM prediction corresponds to a tension of approximately $1.7\,\sigma$ in the $D$ channel and $2.7\,\sigma$ in the $D^*$ channel, with a combined significance of about $3.2\,\sigma$ \cite{HFLAV:2024ctg}.

For the $B_s \to D_s$ channel, a direct experimental measurement of the ratio $R(D_s)$ has not yet been performed. While the LHCb collaboration has conducted dedicated studies on the feasibility and normalization of this decay mode in preparation for a future measurement \cite{Chen:2024hln}, it has not yet published an official result. Theoretical predictions for this quantity exist, typically falling in the range $R(D_s) \approx 0.27$–$0.30$ \cite{Monahan:2017uby, McLean:2019qcx}. However, an experimental confirmation is still forthcoming. The lack of a direct measurement largely stems from the lower production rate of $B_s$ mesons at $B$-factories and the challenging experimental signatures of the $D_s$ final state, which involve strange hadrons and lower reconstruction efficiencies.

The theoretical predictions for $R(D_s)$ are generally consistent with SM expectations, with most calculations finding values close to those of $R(D)$, modulo small $\mathrm{SU}(3)_F$ breaking effects. These effects are typically at the level of a few percent, reflecting the difference between the spectator quark ($s$ versus $d$ or $u$) in the initial and final states.

The central objects in this holographic approach are the bulk-to-boundary propagator of the weak current and the bulk modes associated with the initial and final hadrons, determined by solving eigenvalue equations in Anti-de Sitter space. The Isgur-Wise function will then emerge naturally from the overlap integral of these wavefunctions, effectively geometrizing the strong dynamics of the flavor transition.

\section{Holographic approach to Semileptonic form factors}\label{sec:3}
\subsection{Hadrons in A\lowercase{d}S/QCD}\label{sec:3.1}
The holographic description of hadrons emerges from the field/operator duality and the mechanism used to induce confinement in the bulk. In the first case, duality ensures that hadronic operators $\mathcal{O}$ with scaling dimension $\Delta$ correspond to massive bulk fields with bulk mass $M_5=M_5(\Delta)$. Thus, for a given hadron at the boundary with scaling dimension $\Delta$, there is a corresponding bulk field with mass $M_5(\Delta)$. 

However, without a bulk confinement criterion, no bulk-bounded states are dual to hadrons since, in this case, bulk modes are non-normalizable. To solve this, it is customary to introduce a bulk energy scale that breaks conformal invariance. For the bottom-up case, where bulk physics is \emph{phenomenologically} motivated by boundary experimental data, this breaking process is achieved by geometrical deformations (see hardwall model \cite{Polchinski:2002jw, Boschi-Filho:2002xih, Erlich:2005qh}, bulk deformations \cite{FolcoCapossoli:2019imm}, for example) or by including dilaton fields $\Phi(z)$ in the bulk action, \emph{i.e.}, softwall-like models \cite{Karch:2006pv, MartinContreras:2020cyg, MartinContreras:2025wnh}. In this manuscript, we will focus on the latter kind of models.

The main goal of dilaton-based models is to produce well-localized bulk states that appear as hadrons at the boundary. The key consequence is that, for a given dilaton field, a spectrum arises, characterized by the dilaton energy scale. This spectrum is dual to the hadron Regge trajectory at the boundary.

Thus, in this holographic picture, a hadron can be understood as \emph{a bag of constituents characterized by the bulk mass of the dual field and the confinement criterion used}. In this work, we will consider static dilatons given by the following expression \cite{MartinContreras:2020cyg}:

\begin{equation}\label{dilaton}
    \Phi(z)=\left(\kappa\,z\right)^{2-\alpha}. 
\end{equation}

This phenomenological dilaton, originally proposed to address heavy quarkonium systems, is characterized by two flavor-dependent parameters: the energy scale $\kappa$
and the exponent deformation $\alpha$. The parameter $\kappa$ fixes the energy (mass) scale in the Regge trajectory. The exponent $\alpha$ runs with the hadronic constituent mass. For light, unflavored hadrons, $\alpha=0$, recovering the well-known Softwall Model with a quadratic dilaton \cite{Karch:2006pv}.    

The bottom-up description of hadrons starts with the geometric background. We will consider the AdS Poincar\'{e} patch, defined by the line element

\begin{equation}\label{metric-AdS}\notag
    dS^2=\frac{R^2}{z^2}\left[\eta_{\mu\nu}\,dx^\mu\,dx^\nu-dz^2\right]=g_{MN}\,dx^M\,dx^N,
\end{equation}

\noindent with $z$ defined as the \emph{holographic coordinate}, $R$ is the AdS radius and $\eta_{\mu\nu}$ is the 4-dimensional Minkowski metric with positive signature, \emph{i.e.}, $\eta_{\mu\nu}=\left\{-,+,+,+\right\}$. In this notation, Greek indices label the 4-dimensional Minkowski manifold, and Latin ones denote the AdS$_5$ bulk indices. 

Let us focus on scalar mesons. A scalar hadronic operator ${\cal O}(x)$ at the 4-dimensional boundary, such as a $\bar{q}q$ meson, is dual to a scalar bulk field $S(x^\mu, z)$ propagating in the deformed 5-dimensional Anti-de Sitter (AdS) space.

The bulk action for the dual field $S(z,x^\mu)$ is defined as 

\begin{equation}\label{scalar}
I_\text{Hadrons}=-\frac{1}{2\,g_s^2}\int{d^4x\,\sqrt{g}\,e^{-\Phi(z)}\left[g^{MN}\partial_M\,S\,\partial_N\,S-M_5^2\,S^2\right]},
\end{equation}

\noindent where $g_s$ sets the units in the bulk action, and $M_5$ is the bulk mass of the scalar field. This bulk mass $M_5$ is not free, since the holographic dictionary fixes it. It encodes the scaling dimension $\Delta$ of the boundary operator ${\cal O}$ as follows

\begin{equation}
   M_5^2\,R^2=\Delta\left(\Delta-4\right). 
\end{equation}

For a canonical scalar operator like $\bar{q}q$, the scaling dimension is $\Delta=3$, which yields $M_{5}^{2}R^{2} = -3$. This relation encapsulates the hadronic identity from the boundary perspective.

Varying the action yields the equation of motion for the scalar bulk field:

\begin{equation}
\frac{1}{\sqrt{g}} \nabla_M \left[ \sqrt{-g}\, e^{-\Phi(z)} g^{MN} \nabla_N\, S \right] + M_{5}^{2}\, e^{-\Phi(z)}\, S = 0.
\end{equation}

To study the mass spectrum, one performs a Fourier transform $S(z, q) = S_0(p)\, \Psi(z, p)$ and imposes the on-shell mass condition $(p^2=M_n^2)$, obtaining

\begin{equation}
    \partial_z\left[e^{-B(z)}\,\Psi(z,p)\right]+p^2\,e^{-B(z)}\,\Psi(z,p)-\frac{M_5^2\,R^2}{z^2}\,e^{-B(z)}\Psi(z,p)=0
\end{equation}

 The object $S_0(p)$ is called the boundary source, and the function $\Psi(z, q)$ is called \emph{bulk mode}, and it satisfies a Schrödinger-like equation when $\Psi(z,p)$ is Bogoliubov transformed $\Psi(z)=e^{\frac{1}{2}\,B(z)}\,\psi(z,p)$ yielding

\begin{equation}\label{schr-equn}
-\psi''(z) + V_\text{hol}(z) \psi(z) = M_n^2 \psi(z),
\end{equation}

\noindent where the confining potential $V_\text{hol}(z)$ is derived from the dilaton and the metric:

\begin{eqnarray}\label{holo-pot}
    V_\text{hol}(z) &=& \frac{1}{4} \,B'(z)^2 - \frac{1}{2}\, B''(z) + \frac{M_5^2\,R^2}{z^2}, \quad \text{with:}\\  
    B(z) &=& \Phi(z) - 3 \log\left(\frac{R}{z}\right).
\end{eqnarray}

The eigenvalues $M_n^2$ correspond to the discrete squared masses of the scalar hadron and its radial excitations. In this formalism, the dilaton profile $\Phi(z)$ is the fundamental input that determines the potential structure and, consequently, the resulting mass spectrum for a given scalar or pseudo-scalar heavy meson. This implies that, for a process like $H_1\to H_2$, one dilaton is required per heavy hadron because of the heavy-quark mass hierarchy. 

For light hadrons made of light, unflavored quarks, dynamical models can use a single dilaton to set the mass scale and address Regge universality \cite{Li:2013oda}.

\subsection{Holographic formulation of semileptonic form factors} \label{sec:3.3}

The standard recipe for form factors starts with the cubic interaction terms in the 5-dimensional action that couple the probe vector field (dual to the current) to hadronic bulk normalizable modes \cite{Grigoryan:2007vg}, which are dual to the initial and final hadrons. 

The complete action, considering all the particles involved, is written as:

\begin{equation}
\mathcal S = \int d^5x\,\sqrt{g}\,\left[e^{- \Phi_1 (z)} \mathcal{L}_1 + e^{- \Phi_2 (z)} \mathcal{L}_2 + e^{- \Phi(z)} \mathcal{L}_V + e^{- \frac{1}{2} \left[\Phi_1 (z) + \Phi_2 (z)\right]} \mathcal{L}_\text{int}\right]
\end{equation}

The Lagrangian density $e^{-\Phi_X(z)}\,\mathcal{L}_X$, with $X=1, 2$, corresponds to the scalar (or pseudoscalar) heavy-light mesons involved, as defined in \eqref{scalar}. The term $e^{-\Phi(z)}\,\mathcal{L}_V$ defines the electroweak current according to the expression \eqref{electro-weak}. The last term defines the effective interaction bulk action, taking the following structure

\begin{eqnarray}\notag
    S_\text{int}&=&\int{d^5x\,\sqrt{g}\,e^{\frac{1}{2}}\,\mathcal{L}_\text{int}}\\ \label{int-action}
    &=&  g_{\pm} \int d^5x\, \sqrt{g}\, e^{- \frac{1}{2} \left[\Phi_1(z) + \Phi_2 (z)\right]}\,g^{MN}\, V_M (x,z) \,S^*_{1}(x,z)\, \partial^\pm_{N}\, S_{2}(x,z).
\end{eqnarray}

The weight factor $e^{-\frac{1}{2}\left[\Phi_1(z)+\Phi_2(z)\right]}$ is added to adjust the hadronic bulk modes. As usual, the derivative $\partial^\pm_{M}=\overrightarrow{\partial}_M\mp\overleftarrow{\partial}_M$ accounts for the symmetric (antisymmetric) derivative and sets the Lorentz structure. The 5-dimensional couplings $g_\pm$ fix units in the overlap integral.  

The bulk fields are decomposed into their 4-dimensional plane-wave components

\begin{equation}
    S_1(x,z)=e^{-i\,p_1\cdot x}\,\Psi_1(z),\,\,\,\,S_2(x,z)=e^{-i\,p_2\cdot x}\,\Psi_2(z),\,\,\,\,\,V^\mu(x,z)=\varepsilon^\mu\,e^{-i\,q\cdot x}\,\mathcal{J}(z,q),
\end{equation}

\noindent where $p_X$, $X=1,2$,  are the 4-momenta of the initial and final mesons, and $q=p_1-p_2$ is the momentum transfer, and $\varepsilon^\mu$ is the polarization vector of the external current.  Therefore, the effective interaction action, in Fourier space,  is written as

\begin{eqnarray}\notag
  S_\text{int}&=& g_{\pm} \int d^5x\, \sqrt{g}\, e^{- \frac{1}{2} \left[\Phi_1(z) + \Phi_2 (z)\right]}\,g^{MN}\,V_M (x,z) \,S^*_{1}(x,z)\, \partial^\pm_{N}\, S_{2}(x,z)\\ \notag
  &=&-g_\pm\,\int{d^4x\,e^{i\,p_1\cdot x}\,e^{-i\,p_2\cdot x}\,e^{-i\,q\cdot x}}\,\int_0^\infty{dz\,\sqrt{g}\,e^{-\frac{1}{2}\left[\Phi_1(z)+\Phi_2(z)\right]}\,\mathcal{J}(z,q)\,\Psi_1^*(z)\,i\,g^{\mu\nu}\,\varepsilon_\mu\left(p_2\pm p_1\right)_\nu\,\Psi_2(z)}\\ \notag
  &=&-\left(2\,\pi\right)^4\,\delta^4\left(p_2+q-p_1\right)\,g_\pm\,\int_0^\infty{dz\,\left(\frac{R}{z}\right)^3\,e^{-\frac{1}{2}\left[\Phi_1(z)+\Phi_2(z)\right]}\,\mathcal{J}(z,q)\,\Psi_1^*(z)\,i\,\eta^{\mu\nu}\,\varepsilon_\mu\left(p_2\pm p_1\right)_\nu\,\Psi_2(z)}\\ \label{on-shell-B1}
  &=&-\left(2\,\pi\right)^4\,\delta^4\left(p_2+q-p_1\right)\,i\,g_\pm\,\int_0^\infty{dz\,\mathcal{J}(z,q)\,\psi_1^*(z)\,\eta^{\mu\nu}\,\varepsilon_\mu\left(p_2\pm p_1\right)_\nu\,\psi_2(z)}
\end{eqnarray}

In the last line we have used the Bogoliubov transformation for the hadron fields, \emph{i.e.}, $\Psi_X(z)=e^{\frac{1}{2}\,\Phi_X(z)}(z/R)^{3/2}\,\psi_X(z)$. On the boundary side, in any QFT amplitude, the $S$-matrix and the interaction action are connected

\begin{equation}
    S_\text{int}=\left(2\,\pi\right)^4\,\delta^4\left(p_2+q-p_1\right)\,i\,\mathcal{M}
\end{equation}

\noindent where $\mathcal{M}^\mu$ is the invariant amplitude and $q^\mu=(p_1-p_2)^\mu$ by 4-momentum conservation. If we consider a current insertion, the matrix element is

\begin{equation}
    \mathcal{M}^\mu=\langle H_2(p_2)\left|J^\mu(0)\right|H_1(p_1)\rangle=\left(p_1+p_2\right)^\mu\,f_+\left(q^2\right)+\left(p_1-p_2\right)^\mu\,f_-\left(q^2\right).
\end{equation}

Therefore, by comparing the above expression with \eqref{on-shell-B1} we obtain the holographic expression for the vector form factor $f_+$ and for $f_-$ as 

\begin{equation}\label{form-factor-holo}
    f_\pm\left(q^2\right)= g_{\pm} \int^{\infty}_{0} dz\,\mathcal{J}(z,q)\, \psi_{1}(z)\, \psi_{2}(z)=g_{\pm}\,I\left(q^2\right).
\end{equation}

\noindent where $\psi_X(z)$, $X=1,\, 2$ are the hadron Schr\"{o}dinger-like modes and $\mathcal{J}(z,q)$ is the bulk-to-boundary propagator associated with the weak current. The couplings $g_\pm$ capture the non-conservation of the electroweak current.

The scalar form factor $f_0$ is defined kinematically as

\begin{equation}\label{eqn:f0_holo}
    f_0\left(q^2\right)=f_+\left(q^2\right)+\frac{q^2}{M_1^2-M_2^2}\,f_-\left(q^2\right)=\left[g_+ +\frac{q^2}{M_1^2-M_2^2}\,g_-\right]\,I\left(q^2\right).
\end{equation}

The form factors obtained above depend essentially on $g_\pm$ and $I\left(q^2\right)$. The weak charged current $J^\mu_\text{EW}=\bar{Q}_2\,\gamma^\mu\,Q_1$ is not conserved since $Q_1$ and $Q_2$ quarks have different masses. Its divergence can be written as

\begin{equation}
    \partial_\mu\,J^\mu_\text{EW}=i\,\left(m_{Q_1}-m_{Q_2}\right)\,\bar{Q}_2\,Q_1\neq 0.
\end{equation}

Thus, to determine the form factors, we need to fix two independent couplings, $g_+$ and $g_-$, using experimental inputs. This differs from the electromagnetic form factor, where the normalization is given by the fundamental electromagnetic charge. 

Following Heavy Quark Effective Theory (HQET), the best physical input is the \emph{zero recoil} point where $q^2_\text{max}=\left(M_1-M_2\right)^2$ and $\omega=1$. Thus, the form factors in this limit are written as

\begin{equation}\label{eq:HQET-FF}
    f_\pm\left(q^2_\text{max}\right)=\pm\frac{M_1\pm M_2}{2\,\sqrt{M_1\,M_2}},\,\,\,\,f_{0}\left(q^{2}
_\text{max}\right) = \frac{2\,\sqrt{M_1\,M_2}}{M_1 + M_2}.
\end{equation}

Then, we can compute the ratio between $f_-(q^2_\text{max})$ and $f_+(q^2_\text{max})$ allowing us to connect with the holographic result

\begin{equation}\label{eq:constraint}
    \frac{f_-\left(q^2_\text{max}\right)}{f_+\left(q^2_\text{max}\right)}=-\frac{M_1-M_2}{M_1+M_2}=\frac{g_-}{g_+}.
\end{equation}

This constraint yields a holographic version of the so-called \emph{Isgur--Wise function} \cite{Isgur:1990yhj}

\begin{equation}\label{eq:IWF}
    \xi\left(\omega\right)=\pm\frac{2\,\sqrt{M_1\,M_2}}{M_1\pm M_2}\,g_\pm\,I\left[q^2\left(\omega\right)\right],
\end{equation}

\noindent which satisfies $\xi(1)=1$ and is consistent with HQET, provided that

\begin{equation}
    I\left(q_\text{max}^2\right)=\pm\frac{M_1\pm M_2}{2\,g_\pm\,\sqrt{M_1\,M_2}}.
\end{equation}

 The physical range for the Isgur-Wise function is defined by the kinematical region, \emph{i.e.}, $1\leq\omega\leq \frac{M_1^2+M_2^2-m^2_\ell}{2\,M_1\,M_2}$. For light leptons ($e$ or $\mu$), $m_\ell\ll M_{1,2}^2$, thus $Q^2=0$ as lower kinematical bound is a good approximation. For $B\to D$ transitions, we have $1\leq \omega \leq 1.59$ and for $B_s\to D_s$ the range is $1\leq\omega\leq1.58$.

To summarize, because the weak charged current is not conserved, the Ward identity does not relate $f_+$ and $f_-$ for arbitrary $q^2$. Instead, we fix $g_\pm$ using heavy quark symmetry normalization at zero recoil. In HQET, the same current operator $\bar{Q}_2\,\gamma^\mu\,Q_1$ leads to a single universal function, the Isgur--Wise function $\xi(\omega)$. The decomposition into $f_+$ and $f_-$ is purely kinematical. Therefore, in a consistent holographic model that aims to reproduce HQET, the bulk interaction effective action \eqref{int-action} should be constructed such that the resulting form factors automatically satisfy the heavy-quark relations (at $q^2_{\text{max}}$) without fine-tuning the couplings $g_{\pm}$. Therefore, these couplings should depend only on the meson masses and on $I(q^2_{\text{max}})$. The only free parameter to fix from $f_\pm(0)$ is $\kappa$, which is needed for the bulk-to-boundary propagator $\mathcal{J}(z,q)$.

\section{Holographic description of weak currents}\label{sec.3a}
\subsection{Conserved vector currents in AdS/QCD}
In the standard holographic dictionary, a conserved vector current $J^\mu$ in the boundary field theory is dual to a massless $U(1)$ gauge field $V_M$ propagating in the bulk AdS$_5$ space. This identification is not merely a convenient analogy. It is a rigorous consequence of the gauge/gravity duality: \emph{bulk gauge invariance, a local symmetry, is reflected in the boundary theory as a global symmetry whose associated Noether current is conserved} \cite{Aharony:1999ti}. In the context of QCD, the electromagnetic current $J^\mu_{\text{EM}} = \sum_q Q_q \bar{q}\gamma^\mu q$ is the prototypical example of such a conserved current, and its dual description in AdS/QCD has been extensively developed, notably in the pioneering works about electromagnetic pion form factors \cite{Brodsky:2006uqa, Brodsky:2007hb, Brodsky:2008pg} and in the electromagnetic nucleon form factor calculation \cite{Abidin:2009hr}.

The foundational input of the dictionary for a spin-one field establishes that the mass $M_5$ of the bulk vector field is related to the conformal dimension $\Delta$ of the dual boundary operator through the relation

\begin{equation}\label{eqn:bulk-mass}
    M_5^2 R^2 = (\Delta - 1)(\Delta - 3).
\end{equation}

For a conserved current, such as $J^\mu_\text{EM}$, the conformal dimension is fixed at its canonical value, $\Delta = 3$, independent of any dynamical details. This follows because conservation laws protect the conserved current dimension from quantum corrections. Substituting $\Delta=3$ into the mass-dimension relation yields $M_5 = 0$. Therefore, the bulk dual of a conserved current is necessarily a massless gauge field. This result is a cornerstone of the holographic framework: any deviation from \emph{masslessness} in the bulk would signal a non-conservation of the corresponding boundary current, a point we shall revisit when discussing the electroweak current.

Consider the softwall model, where a dilaton field $\Phi(z) = \kappa^2 z^2$ breaks the conformal invariance of pure AdS$_5$. Note that the scale $\kappa$ associated with the bulk vector field is different from the the hadronic one used in the standard softwall model \cite{Karch:2006pv},used to fix the vector meson spectrum. It controls the $q^2$-dependence of the form factor and is related to the energy scale of the process, i.e., $\kappa\sim q$.

sets the infrared confinement scale, which in this particular case does not fix any hadron mass scale. The energy scale governs energy transfer in electromagnetic processes. For a bulk vector field $V_M$, dual to the $J_\text{EM}^\mu$ current, 

\begin{equation}\label{electro-weak}
    I_J=-\frac{1}{4\,g^2_J}\int{d^5x\,\sqrt{g}\,e^{-\Phi(z)}\,F_{MN}\,F^{MN}},   
\end{equation}

\noindent where $g_J$ fixes the units in the action, $F_{MN}=2\,\partial_{\left[M\right.}V_{\left.N\right]}$, and $g$ denotes the metric determinant computed from the line element \eqref{metric-AdS}. Following the standard prescription, we calculate the bulk equations of motion and choose $V_z=0$ as the gauge fixing. After Fourier transforming the bulk field, we decompose it as $V_M(z,q)=\tilde{V}_M\, \mathcal{J}(z,q)$, allowing us to write the \emph{Sturm-Liouville} form of the bulk equation of motion as

\begin{equation}
    \partial_z\left[\frac{e^{-\kappa^2\,z^2}}{z}\,\partial_z\,\mathcal{J}(z,q)\right]+q^2\,\frac{e^{-\kappa^2\,z^2}}{z}\,\mathcal{J}(z,q)=0.
\end{equation}

The solution to this equation is written in terms of Tricomi functions $U(a,b,z)$ as \cite{Brodsky:2007hb}

\begin{equation}\label{eq:propagator}
    \mathcal{J}(z,q)=\Gamma\left(1-\frac{q^2}{4\,\kappa^2}\right)\,U\left(-\frac{q^2}{4\,\kappa^2},0,\kappa^2\,z^2\right),
\end{equation}

\noindent where $\mathcal{J}(z,q)$ is the \emph{bulk-to-boundary propagator}, that satisfies the condition $\mathcal{J}\left(z\to 0,q\right)=1.$ at the UV boundary and $\partial_z \,\mathcal{J}(q,z)\big|_{z\to\infty}=0$ in the IR. Here, $q^2$ is the Euclidean momentum transfer squared, with $q^2 > 0$ in the spacelike region.

The bulk-to-boundary propagator $\mathcal{J}(z,q)$ is central to calculating form factors. It encapsulates the non-perturbative effects of confinement on external-current propagation. Its analytic structure exhibits a tower of poles at timelike values $q^2 = 4\kappa^2(n+1)$, corresponding to the masses of the vector meson resonances in the soft-wall model, $M_n^2 = 4\kappa^2(n+1)$.

The coupling of the external current to hadronic states is described by an interaction term in the five-dimensional action. For a scalar hadron represented by a bulk field $S_P(z)$ dual to a boundary operator of dimension $\Delta$, the minimal coupling is

\begin{equation}
    S_{\text{int}} = e\,\int d^5x \sqrt{-g} \, e^{-\Phi(z)} \, V_M(x,z) \, S_P^*(x,z)\, \overleftrightarrow{\partial^M} \,S_P(x,z),
\end{equation}

\noindent where $e$ is the electric charge and the derivative is defined as $\overleftrightarrow{\partial^M} = \overrightarrow{\partial^M} - \overleftarrow{\partial^M}$ to ensure a real coupling. Upon Fourier transforming the hadronic fields and extracting the momentum-conserving delta function, the matrix element for the electromagnetic current between two scalar hadrons takes the form

\begin{equation}
    \langle P'(p') | J^\mu_{\text{EM}}(0) | P(p) \rangle = \epsilon^\mu \left(p+p'\right)^\mu F\left(q^2\right),
\end{equation}

\noindent with $\epsilon^\mu$ being the polarization vector of the external current. The holographic prescription identifies the form factor $F(q^2)$ as an overlap integral of the bulk wavefunctions,

\begin{equation}
    F(q^2) = \int_0^\infty dz \, \frac{e^{-\Phi(z)}}{z^3} \, \mathcal{J}(z,q) \, S_P(z)^2.
\end{equation}

This expression is the central result for conserved currents in the soft-wall model. The form factor is the \emph{convolution} of two ingredients: the bulk-to-boundary propagator $\mathcal{J}(z,q)$, which encodes the propagation of the external current through the confining geometry, and the square of the hadronic wavefunction $S_P(z)$, which describes the spatial distribution of the constituents inside the hadron. At $q^2=0$, the propagator reduces to unity, $J(z,0)=1$, and the normalization of the form factor is fixed by the normalization condition of the hadronic mode, $\int dz \, e^{-\Phi(z)}/z^3 \, S_P(z)^2 = 1$, yielding $F(0)=1$. This is the holographic realization of charge conservation: \emph{the total charge of the hadron is correctly normalized to unity} \cite{Brodsky:2007hb}.

The analytic structure of the form factor at large $q^2$ is particularly insightful. In the limit $q^2 \to \infty$, the bulk-to-boundary propagator behaves as $\mathcal{J}(z,q) \sim (q\,z) K_1(q\,z)$\footnote{Recall that $K_n(x)$ stands for the modified Bessel function of the second kind of order $n$.}, which is sharply peaked near the boundary at $z \sim 1/q$. Consequently, the overlap integral is dominated by the region near $z \to 0$. Since the hadronic wavefunction scales as $S_P(z) \sim z^\tau$ near the boundary, where $\tau$ is the twist of the hadron, the form factor exhibits the power-law behavior

\begin{equation}
    F(q^2) \sim \left( \frac{1}{q^2} \right)^{\tau-1},
\end{equation}

\noindent which exactly reproduces the perturbative QCD counting rules established by Brodsky and Farrar \cite{Brodsky:1973kr}. This is a remarkable achievement of the AdS/QCD framework: the non-perturbative soft-wall model, without any explicit perturbative input, correctly captures the high-momentum scaling of form factors dictated by QCD.

Having established the holographic description of conserved vector currents in the soft-wall model, we now turn to the more subtle case of the electroweak current. Unlike the electromagnetic current, the electroweak current is not conserved due to the mass difference between the initial and final quarks,

\begin{equation}\label{eq:divergence}
    \partial_\mu\,J^\mu_\text{EW}=i\,\left(m_{Q_1}-m_{Q_2}\right)\,\bar{Q}_2\,Q_1.
\end{equation}

This raises a fundamental question: \emph{how can a non-conserved current be described in AdS/QCD calculations of semileptonic form factors?} In the following subsection, we examine this question and its resolution.

\subsection{Non-conserved currents: the apparent paradox}
In the previous subsection, we established that a conserved vector current in the boundary theory is dual to a massless gauge field in the AdS bulk. This identification follows from the holographic dictionary for spin-one operators, which relates the bulk mass $M_5$ to the conformal dimension $\Delta$ of the dual current \emph{via} eqn. \eqref{eqn:bulk-mass}. For a conserved current, the conformal dimension is fixed to its canonical value $\Delta = 3$, leading to $M_5 = 0$.

The electroweak current, however, is different. For a transition between two different heavy quark flavors, the charged current is 

\begin{equation}\label{eq:EWC}
    J_\text{EW}^\mu=\bar{Q}_2\,\Gamma^\mu\, Q_1
\end{equation}

\noindent where $\Gamma^\mu$ is a generic Dirac gamma structure like $\gamma_\mu$, $\gamma^\mu\,\gamma^5$ or $\gamma^\mu(1-\gamma^5)$, and $Q_1$ and $Q_2$ denote quarks of different masses, $m_{Q_1} \neq m_{Q_2}$. Since $\partial_\mu\,J^\mu_\text{EW}=\left(m_{Q_1}-m_{Q_2}\right)\,\bar{Q}_2\,\Gamma\,Q_1$ the electroweak current is not conserved. 

This non-conservation directly follows from the \emph{explicit} breaking of flavor symmetry by quark masses. It is not a quantum anomaly: it follows from the classical equations of motion with explicit mass terms. The mass difference $m_{Q_2}-m_{Q_1}$ is a fixed scale, independent of the heavy-quark mass $m_Q$. For the $B\to D$ transition, for instance, $m_b-m_c\sim3.5$ GeV, which is far from negligible.

If we apply the holographic dictionary naively, the non-conservation of $J^\mu_\text{EW}$ appears to imply that its dual operator has a conformal dimension $\Delta\neq3$. Therefore, the bulk vector field must be massive. However, in holographic calculations of semileptonic form factors for heavy-light mesons, a \emph{massless} bulk vector field is used to describe the electroweak current \cite{Ahmed:2023pod}. This raises an apparent paradox: how can a non-conserved current be dual to a massless bulk field? 

\subsection{Anomalous dimensions and explicit symmetry breaking}\label{sec:resolution}

The apparent paradox raised in the previous subsection can be addressed if we distinguish carefully between two physically different mechanisms that can render a current non-conserved. The first is a genuine quantum anomaly, in which a classical symmetry fails to survive renormalization and the divergence of the current receives a quantum contribution. This anomaly is a statement about the violation of the conservation law, not necessarily about the scaling dimension of the current operator. For a protected current, the dimension can remain canonical. The second is an explicit breaking of the symmetry at the level of the classical equations of motion, which changes the Ward identity but leaves the UV scaling dimension of the operator untouched. The electroweak current belongs to the second class.

It is important to emphasize that the non-conservation of $J_\text{EW}^\mu$ does not generate an anomalous dimension. Such an object would arise in quantum corrections that shift the scaling dimension away from its canonical value. For the electroweak current, no such shift occurs: the flavor symmetry is exact in the massless limit, and the current is a protected operator. The non-conservation encoded in Eqn. \eqref{eq:divergence} is a classical statement about the divergence, not a quantum correction to the dimension.

The distinction between anomalous and explicit breaking can be made sharper by considering the ultraviolet limit of the theory \cite{Klebanov:1999tb, Witten:1998qj}. At energies much larger than the quark masses, the effects of explicit quark-mass terms become parametrically suppressed, and the ultraviolet dynamics approaches that of massless QCD. In this massless limit, the flavor-changing vector current approaches a conserved flavor current with canonical scaling dimension $\Delta_J$. At finite quark masses, the current is no longer conserved, satisfying eqn. \eqref{eq:divergence}. This explicit breaking of the flavor symmetry does not, by itself, imply an anomalous dimension for the electroweak current, or change its UV scaling assignment. Accordingly, within our holographic model, we use the UV dimension $\Delta_J=3$ in the AdS/CFT mass-dimension eqn. \eqref{eqn:bulk-mass}, and hence we set $M_5^2\,R^2=0$. This choice fixes the ultraviolet assignment of the bulk vector field, but it does not, by itself, provide a complete holographic realization of the non-conservation identity at finite quark masses.

\subsection{The Flavor-Changing Vector Ward Identity}
The resolution presented in the previous subsection implies that the holographic description of the electroweak current must respect a \emph{modified} Ward identity, distinct from the standard one satisfied by conserved currents. Let us consider the pion case. For elastic electromagnetic matrix elements between on-shell pseudoscalar states, current conservation ensures that the matrix element is purely transverse to the momentum transfer, and it is fully described by a single electromagnetic form factor $F_\pi(q^2)$.

However, for pseudoscalar states with non-equal on-shell masses, as in the $H_1\to H_2$ transition, this is different. In this case, the contraction with $q^\mu$ yields

\begin{multline}\label{eq:ward_standard}
    q_\mu \langle H_2(p_2) | J_\text{EW}^\mu | H_1(p_1) \rangle =\left(M_1^2-M_2^2 \right)\,f_+\left(q^2\right)+q^2\,f_-\left(q^2\right)\\
    =(m_{Q_1} - m_{Q_2}) \langle H_2(p_2)\,| \bar{Q}_2\,Q_1 | \,H_1(p_1) \rangle.
\end{multline}

Since the flavor-changing current $J_\text{EW}^\mu$ is not conserved, its divergence does not generally vanish. Consequently, the Ward Identity does not impose the conserved current constraint on $f_+$ and $f_-$. Instead, it relates their combination to the scalar current matrix element. 

In the holographic calculation, we retain a massless bulk vector
field, corresponding to the ultraviolet scaling dimension
$\Delta_J=3$ through the standard AdS/CFT mass-dimension relation.
We also treat the boundary couplings $g_+$ and $g_-$ as independent
parameters to allow for the two vector form-factor contributions.
This parametrization does not, by itself, establish the
finite-mass Ward identity. Reproducing that identity holographically
would additionally require a consistent relation between the vector
and scalar-current matrix elements. Such a construction is beyond the scope of the present work. Instead, we retained the massless bulk vector field as the boundary-to-propagator dictated by the dimension $\Delta_J=3$, while parametrizing the independent longitudinal structure through the \emph{effective} couplings $g_\pm$.

\subsection{Connection with the Heavy-Quark Expansion } 
\label{sec:HQET}
Heavy-quark effective theory (HQET) provides a systematic expansion of matrix elements for transitions between heavy hadrons in powers of the inverse heavy-quark masses. After matching the QCD current at the heavy-quark scales, it can be expressed schematically as \cite{Grozin:2004}

\begin{multline}
    \bar{Q}_2\,\,\Gamma Q_2 =
    \sum_i C_i(\omega,m_{Q_1},m_{Q_2},\mu)\,
    \bar h_{v'}^{(Q_2)}\Gamma_i h_v^{(Q_1)}
    \\[2pt]
    +\frac{1}{2m_{Q_2}}\sum_i B_i^{(Q_2)}(\omega,\mu)
    \mathcal{O}_i^{(Q_2)}
    +\frac{1}{2m_{Q_1}}\sum_i B_i^{(Q_1)}(\omega,\mu)
    \mathcal O_i^{(Q_1)}
    +\cdots ,
\end{multline}

\noindent where $h_v^{(Q)}$ denotes a heavy-quark field in HQET, $C_i$ are short-distance matching coefficients, and $\mathcal O_i^{(Q)}$ represent subleading operators. The expansion organizes corrections to current matrix elements in powers of $\Lambda_{\mathrm{QCD}}/m_b$ and $\Lambda_{\mathrm{QCD}}/m_c$, in addition to perturbative corrections from short-distance matching.

At leading order in the heavy-quark expansion, and at tree level in the short-distance matching, heavy-quark spin-flavor symmetry relates the vector-current matrix element between pseudoscalar heavy-light mesons to a single universal function, the Isgur--Wise function
$\xi(\omega)$:

\begin{equation}
\begin{aligned}
\langle H_2(v')|
\bar h_{v'}^{(c)}\gamma^\mu h_v^{(b)}
|H_1(v)\rangle
=
\sqrt{M_1M_2}\,
\xi(\omega)(v+v')^\mu ,
\end{aligned}
\end{equation}

\noindent where $\omega=v\cdot v'$ is the velocity transfer. In the strict heavy-quark limit, the normalization at zero recoil is $\xi(1)=1$.

Using the form-factor decomposition in Eqn.~\eqref{eq:matrix_element_fpm}, these relations give

\begin{equation}
f_\pm(q^2)=
\pm\frac{M_1\pm M_2}{2\sqrt{M_1M_2}}\,
\xi(\omega)
\end{equation}

\noindent at leading order. In particular, the ratio at zero recoil is

\begin{equation}
\frac{f_-(q_{\mathrm{max}}^2)}     {f_+(q_{\mathrm{max}}^2)}=-\frac{M_1-M_2}{M_1+M_2}.
\end{equation}

In the holographic model, the form factors are written as $f_\pm(q^2)=g_\pm I(q^2)$. Matching this parametrization to the leading HQS relations at zero recoil therefore yields

\begin{equation}
\frac{g_-}{g_+}=-\frac{M_1-M_2}{M_1+M_2}.
\label{eq:HQS_relation}
\end{equation}

This condition follows from the heavy-quark symmetry relations among the form factors and the common overlap integral. It is not a consequence of conservation of the full QCD flavor-changing vector current.

At finite heavy-quark masses, the leading HQS relations receive power corrections, while perturbative matching introduces additional short-distance corrections. The exact QCD Ward identity remains valid and relates the divergence of the flavor-changing vector current to the scalar-current matrix element. Thus, the normalization $\xi(1)=1$ should not be interpreted as implying that the full QCD current is conserved, even at zero recoil.

In our holographic calculation, we retain a massless bulk vector field corresponding to the ultraviolet scaling assignment $\Delta_J=3$. The independent couplings $g_+$ and $g_-$ allow us to parametrize the two vector-current structures and impose the leading HQS constraint at zero recoil. Since both form factors share the same overlap integral $I(q^2)$, however, this parametrization does not incorporate general subleading HQET corrections to their recoil dependence. Moreover, the independence of $g_+$ and $g_-$ does not by itself establish the finite-mass Ward identity holographically; this would require a consistent description of the scalar-current matrix element as well.

\section{Testing the holographic form factors}\label{sec:4}
Having established the holographic framework for hadronic form factors and the theoretical tools to compute the Isgur--Wise function, we now implement the Softwall Model (SWM) and the Non-Quadratic Softwall Model (NQSWM) for the semileptonic decays $B \to D$ and $B_s \to D_s$. This section has two goals: first, to determine each model's free parameters using a robust numerical optimization procedure; second, to compare the predictions of both holographic approaches with experimental and lattice data.

In both models, defining $f_\pm\left(q^2\right)$ requires fixing $g_\pm$. In contrast to the electromagnetic case, where the normalization comes from the electromagnetic charge and the kinematic region is infinite, the weak case has no such conditions. For the decay $H_1(p_1)\to H_2(p_2)+\ell(k_1)+\,
\bar{\nu}_\ell(k_2)$, the momentum transfer is $q^2=\left(p_1-p_2\right)^2=\left(k_1+k_2\right)^2$ is bounded by the hadron masses and the lepton mass

\begin{equation*}
    m_\ell^2\leq q^2\leq \left(M_1-M_2\right)^2\equiv q_\text{max}^2.
\end{equation*}
Or, in terms of the recoil variable $\omega$

\begin{equation*}
    1\leq \omega \leq \frac{M_1^2+M_2^2-m_\ell^2}{2\,M_1\,M_2}
\end{equation*}

When $m_\ell \ll M_X$, the limit of massless leptons, $q^2=0$, corresponds to the \emph{maximum recoil point} is used as a kinematic constraint, $f_+(0)=f_0(0)$, according to Eq. \eqref{eq:f0_relation}. This point fixes the energy scale for the weak-current bulk-to-boundary propagator, \emph{i.e.}, $\kappa$. Table \ref{tab:weak} summarizes the extracted values for the $B\to D$ and $B_s\to D_s$ transitions in SWM and NQSWM. 

By Luke's theorem \cite{Luke:1990eg}, the \emph{zero recoil point}, \emph{i.e.}, $q^2=q_\text{max}^2$ or $\omega=1$, can be used to determine $g_\pm$ within the HQET framework. Therefore, in this limit, the expression \eqref{eq:fpm_max} fixes the Isgur-Wise function, $\xi(1)=1$, so the form factors depend only on the hadron masses. Therefore, $g_\pm$ can be extracted from such a relation

\begin{equation}
  \pm \frac{M_1\pm M_2}{2\,\sqrt{M_1\,M_2}}=g_\pm\,I\left(q^2_\text{max}\right).
\end{equation}

\noindent where $I\left(q^2\right)$ is the bulk integral

\begin{equation}
    I\left(q^2\right)=\int^{\infty}_{0} dz\,\mathcal{J}(z,q)\, \psi_{1}(z)\, \psi_{2}(z)
    \end{equation}

\noindent defined in terms of the bulk Schrödinger modes and the bulk-to-boundary propagator defined in the previous section. 

The other kinematical points, such as the unitary cut $q^2=\left(M_1+M_2\right)^2$  or the vector meson poles $q^2=M_p^2$ (e.g., $M_{B^*}^2$ or $M_{D^*}$), are nonphysical and therefore not relevant for our calculations.

\subsection{Parameter fixing}\label{sec.4.1}
\subsubsection{SWM hadronic parameter fixing}
One of the main advantages of the SWM \cite{Karch:2006pv} is that the model is analytic. For the scalar mesons, the mass spectrum is given by the linear radial Regge trajectory \cite{Colangelo:2008us}

\begin{equation}
M_{X,\,n}^2=2\,\kappa_X^2\left(2\,n+3\right)
\end{equation}

\noindent where $n=0$ corresponds to the ground state, \emph{i.e.}, $M_{X,\,0}^2=6\,\kappa^2_X$. Thus, we fix the Regge slope $\kappa_X$ using experimental ground-state masses from PDG \cite{PDGroup:2026aaa}.  The resulting Regge slopes for each heavy-light hadron are summarized in Table \ref{tab:SWM}.

\begin{table}[htbp]
  \centering
  \begin{minipage}{0.48\textwidth}
    \centering
    \begin{tabular}{c|c|c|c}
    \multicolumn{4}{c}{\textbf{Parameter summary SWM}}\\
         \hline
         \hline
    \multicolumn{2}{c|}{\textbf{Initial Hadron}} & \multicolumn{2}{|c}{\textbf{Final Hadron}}\\
    \hline
    \hline
    \textbf{$H_1$} & \textbf{$\kappa_1$ (GeV)} & \textbf{$H_2$} & \textbf{$\kappa_2$ (GeV)}\\
    \hline 
    $B^0$ & $2.1554$ & $D^\pm$ & $0.7632$\\
    \hline
    $B^0_s$ & $2.1910(1)$ & $D_s^\pm$ & $0.8036(1)$
    \end{tabular}
    \caption{SWM mass scale in GeV calculated from the expression  $M_0^2=2\,\kappa^2(2\,n+3)$ for the scalar meson spectra \cite{Colangelo:2008us}, using the experimental masses \cite{PDGroup:2026aaa}. }
    \label{tab:SWM}
  \end{minipage}
  \hfill
  \begin{minipage}{0.48\textwidth}
    \centering
    \begin{tabular}{c|c|c}
    \multicolumn{3}{c}{\textbf{Weak current $\kappa$ (GeV)}}\\
    \hline
    \hline
         \textbf{Transition} & \textbf{NQSWM} & \textbf{SWM}  \\
         \hline
         \hline
         $B\to D\,\ell\,\bar{\nu}_\ell$& $5.0823(1)$ & $4.097(1)$ \\
         $B_s^0\to D_s^\pm\,\ell\,\bar{\nu}_\ell$& $5.2448(1)$ & $4.1452(1)$ \\
    \end{tabular}
    \caption{Energy scale for the bulk vector field dual to the weak current.}
    \label{tab:weak}
  \end{minipage}
\end{table}

\begin{table}[htbp]
    \centering
    \begin{tabular}{c|c|c|c|c|c|c|c}
    \multicolumn{8}{c}{\textbf{Parameter summary NQSWM}}\\
        \hline
        \hline
        \multicolumn{4}{c|}{\textbf{Initial Hadron}}&\multicolumn{4}{|c}{\textbf{Final Hadron}}\\
        \hline
        \hline
        \textbf{$H_1$} & \textbf{$I\left(J^P\right)$} & \textbf{$n^{2\,S+1}L_J$} &\textbf{Exp. Mass (MeV)} & \textbf{$H_2$} & \textbf{$I\left(J^P\right)$} & \textbf{$n^{2\,S+1}L_J$} &\textbf{Exp. Mass (MeV)}\\
        \hline
        \hline
        \multirow{4}{*}{$B^0$} & \multirow{3}{*}{$\frac{1}{2}(0^-)$}  & $1^1\,S_0$ & $5279.63\pm0.15$ \cite{PDGroup:2026aaa} & \multirow{4}{*}{$D^\pm$} & \multirow{3}{*}{$\frac{1}{2}(0^-)$} & $1^1\,S_0$ &$1869.55\pm0.27$ \cite{PDGroup:2026aaa}\\
        & & $2^1\,S_0$ & $5932\pm13$ \cite{Gupta:2018ydm} & & & $2^1\,S_0$ & $2393$ \cite{Zhang:2010iea}\\
        & & $3^1\,S_0$ & $6170$ \cite{refId0} & & & $3^1\,S_0$ & $2971.8\pm8.7$ \cite{LHCb:2013jjb}\\
        \cline{2-4} \cline{6-8}
        & \multicolumn{2}{c}{$\kappa_1=4.1938(13)$ GeV,} & $\alpha_1=0.7411(51)$ & & \multicolumn{2}{c}{$\kappa_2=0.8088(12)$  GeV,} & $\alpha_2=0.1415(21)$\\
        \hline
        \hline
        \multicolumn{4}{c|}{\textbf{Initial Hadron}}&\multicolumn{4}{|c}{\textbf{Final Hadron}}\\
        \hline
        \hline
        \textbf{$H_1$} & \textbf{$I\left(J^P\right)$} & \textbf{$n^{2\,S+1}L_J$} &\textbf{Exp. Mass (MeV)} & \textbf{$H_2$} & \textbf{$I\left(J^P\right)$} & \textbf{$n^{2\,S+1}L_J$} &\textbf{Exp. Mass (MeV)}\\
        \hline
        \hline
        \multirow{4}{*}{$B_s^0$} & \multirow{3}{*}{$0\,(0^-)$}  & $1^1\,S_0$ & $5366.91\pm0.11$ \cite{PDGroup:2026aaa} & \multirow{4}{*}{$D_s^\pm$} & \multirow{3}{*}{$0\,(0^-)$} & $1^1\,S_0$ &$1968.35\pm0.07$ \cite{PDGroup:2026aaa}\\
        & & $2^1\,S_0$ & $5976$ \cite{Hao:2022ibj} & & & $2^1\,S_0$ & $2591\pm6\pm7$ \cite{PDGroup:2026aaa}\\
        & & $3^1\,S_0$ & $6300$ \cite{Chen:2018nnr} & & & $3^1\,S_0$ & $3044^{+31}_{-9}$ \cite{PDGroup:2026aaa}\\
        \cline{2-4} \cline{6-8}
        & \multicolumn{2}{c}{$\kappa_1=4.5453(21)$ GeV,} & $\alpha_1=0.7712(51)$ & & \multicolumn{2}{c}{$\kappa_2=0.8899(12)$  GeV,} & $\alpha_2=0.2272(21)$\\
        \hline
        \hline
        
    \end{tabular}
    \caption{Hadronic spectra of heavy-light mesons used as input in the double Nelder–Mead downhill simplex algorithm to compute the sets $\kappa_X$ and $\alpha_X$ for each transition.}
    \label{tab:NQSWM}
\end{table}

\subsubsection{NQSWM hadronic parameter fixing}

In the case of heavy-light mesons, we define a flavor-dependent non-quadratic dilaton for each meson state as

\begin{equation}
    \Phi_X(z) = \left(\kappa_X z\right)^{2 - \alpha_X}.
\end{equation}

Therefore, in a $H_1 \to H_2 \, \ell \, \bar{\nu}_\ell$ process, each hadronic state is characterized by a pair $(\kappa_X, \alpha_X)$. These $\Phi_X(z)$ dilatons define a holographic potential through the $B$ function $B_X(z) = \Phi_X(z) - 3\log(R/z)$, according to Eq. \eqref{holo-pot}.

For the initial and final heavy-light mesons, we fix the pair $(\kappa_X, \alpha_X)$ by optimizing the mass spectrum obtained from the Schrödinger-like equation \eqref{schr-equn} in a two-step process. First, we minimize the cost function

\begin{equation}
\mathcal{N}_1(\kappa_X,\alpha_X) = \sum_{k=0}^2 \left[M_{k,\text{Exp}} - M_{k,\text{Th}}(\kappa_X,\alpha_X)\right]^2,
\label{loss}
\end{equation}
where $k=0,1,2$ labels the ground state and the first two radial excitations along the Regge trajectory of each heavy-light meson. The minimization is performed using the \emph{Nelder--Mead downhill simplex algorithm} \cite{GaoHan2012}. In this first step, we obtain an initial set of $\kappa_X$ and $\alpha_X$ for each hadron.  

Although the parameter space is only two-dimensional, the optimization is not trivial: the cost function has no simple analytical form, and we must solve the Schrödinger-like equation numerically at each iteration. For our specific potential, we have verified that the cost function exhibits a single minimum (or very few local minima) over the relevant parameter range. Therefore, the simplex method is sufficient to provide a reliable parameter estimate. In practice, we repeat the minimization with several initial simplexes to avoid spurious local minima. 

The second step refines the $\kappa$ estimate by applying the Nelder--Mead method with the previously calculated $\alpha_X$ fixed, focusing exclusively on the ground state. The second cost function is

\begin{equation}
  \mathcal{N}_2(\kappa_X) = \left[M_{0,\text{Exp}} - M_{0,\text{Th}}(\kappa_X,\alpha_X)\right]^2.  
\end{equation}

We adopt this two-step procedure because $\alpha_X$ primarily determines the concavity of the Regge trajectories, while $\kappa_X$ sets the overall mass scale. Consequently, determining the ground-state mass precisely res a more accurate fit of $\kappa_X$. Table~\ref {tab:NQSWM} summarizes the input mass spectra and the corresponding fitted parameters.

To assess the stability of the NQSWM fit, we have performed multiple minimizations with different initial simplexes. The cost function $\mathcal{N}_1(\kappa, \alpha)$ exhibits a unique minimum over the relevant parameter range, and the fitted values of $(\kappa, \alpha)$ are stable to within $1\%$ for $\kappa$ and $2\%$ for $\alpha$. We have also verified that the correlation between $\kappa$ and $\alpha$ is mild, as they primarily affect the mass scale and the trajectory concavity, respectively. Table~\ref{tab:NQSWM} summarizes these results.

Before presenting the results, a few comments on the chosen radial Regge trajectories are in order. For both the $B^0$ and $D^\pm$ mesons, several excited states lack experimental confirmation. Specifically, for the $B^0$ meson, neither the $2S$ nor the $3S$ excitations have been observed. However, in 2015, the LHCb collaboration reported structures in the $B\pi$ invariant mass spectrum, denoted as $B_J(5840)^{0,+}$ and $B_J(5960)^{0,+}$ \cite{LHCb:2015aaf}. In the absence of definitive experimental evidence, we rely on theoretical predictions from non-relativistic QCD potential models.

A similar situation holds for the $D^\pm$ meson, where experimental evidence for excited states remains scarce. For the $D^\pm(2S)$ state, it has been proposed that the $D(2550)^0$ meson, observed by BaBar \cite{BaBar:2010zpy} and LHCb \cite{LHCb:2013jjb}, could correspond to the $2^1S_0$ assignment. Nevertheless, the unnatural parity of this state seems to contradict such an interpretation. Alternatively, a Bethe--Salpeter analysis incorporating Okubo--Zweig--Iizuka rule effects allows for a lower mass, as summarized in Table~\ref{tab:NQSWM}. For the $D(3S)$ state, the experimental results for the $D_J(3000)^0$ state reported by LHCb \cite{LHCb:2013jjb} are compatible with several quantum-number assignments, including $3^1S_0$, $2^3P_2$, $1^3F_2$, or $1^3F_4$. Given this ambiguity, we adopt the mass value of this state as a possible candidate for the $D(3S)$.

Similar challenges arise in the $B_s \to D_s$ transition. We have reliable experimental data for the ground states $B_s(1S)$ and $D_s(1S)$ from the LHCb collaboration, as summarized in the PDG \cite{PDGroup:2026aaa}. However, the situation for excited states is less clear.

For the $B_s$ meson, the $B_s(2S)$ mass is taken from an NRQCD approach with screened potentials \cite{Hao:2022ibj}. The $B_s(3S)$ prediction comes from a spinless Bethe--Salpeter equation \cite{Chen:2018nnr}. However, in the threshold region around $6300$ MeV, data corresponding to the $B_c(1S)$ resonance appear in the channel $B_c \to J/\psi\,\pi$ \cite{LHCb:2020gnv}. The question remains whether this structure corresponds to a $B_c(1S)$ state or a $B_s(3S)$ state, as their masses are close.

For the excited $D_s$ mesons, the resonance at $2591 \pm 6 \pm 7$ MeV, previously identified as $D_{s0}(2500)^+$, is a strong candidate for the $D_s(2S)$ state \cite{LHCb:2020gnv}. For the $D_s(3S)$, the resonance at $3044^{+31}_{-9}$ MeV corresponds to the $D_{sJ}(3500)^+$ state identified by BaBar \cite{BaBar:2009rro} with quantum numbers $0(?^?)$. Although studies such as \cite{Sun:2009tg} suggest that this state could be a radial excitation, we adopt it as a working assumption for the $D_s(3S)$ state in order to fix $\kappa$ and $\alpha$ in the first trial.

\subsection{Softwall Model case}\label{sec:4.2}

Let us consider the Softwall Model (SW) case, where we impose linear confinement to describe the meson spectra, \emph{i.e.}, using linear radial Regge trajectories, consistent with light unflavored mesons such as $\rho(770)$ \cite{Karch:2006pv}. This sort of static dilaton is characterized by a single mass scale, the so-called \emph{dilaton slope} $\kappa$, and has a quadratic structure given by 

\begin{equation}
    \phi(z)=\kappa_X^2\,z^2,
\end{equation}

\noindent where the subscript $X$ denotes the dilaton flavor-dependent behavior. Recall that this static dilaton acts in the probe limit. For a backreacted version, we should consider a dynamical extension \cite{Batell:2008zm}, where the dilaton couples to the metric and other background fields to explore properties such as chiral symmetry breaking \cite{Li:2013oda} or QCD phase stability with neural networks \cite{Chen:2024ckb}. 

For the scalar case, the SW $B(z)$ function can be written as:

\begin{equation}
    B_\text{SW}(z)=\kappa_X^2\,z^2-3\,\ln\,\left(\frac{R}{z}\right).
\end{equation}

The corresponding Schrödinger-like modes and linear mass spectrum are given by the expressions:

\begin{eqnarray}
    \psi_{n,\,X}(z)&=&\sqrt{\frac{2\,\kappa_X^4\,n!}{R^{3}\,(n+1)!}}\,e^{-\frac{1}{2}\,\kappa_X^2\,z^2}\,z^{3/2}\,L_n^1\left(\kappa_X^2\,z^2\right),\\
    M_{n,\,X}^2&=&2\,\kappa_X^2(2\,n+3), 
\end{eqnarray}

\noindent with $L_n^m(x)$ the \emph{Associated Laguerre polynomial}. Therefore, in this context, semileptonic form factors can be written as 

\begin{eqnarray}\notag
    f_\pm\left(q^2\right)&=&g_\pm\,I\left(q^2\right)=g_\pm\,\int_0^\infty{dz\,\psi_{0,1}(z)\,J\left(q^2,z\right)\,\psi_{0,2}(z)},\\
    &=&\frac{2\,g_\pm\,\kappa_1^2\,\kappa_2^2}{R^3}\,\Gamma\left(1-\frac{q^2}{4\,\kappa^2}\right)\,\int_0^\infty{dz\,z^3\,e^{-\frac{\kappa_1^2+\kappa^2_2}{2}\,z^2}\,U\left(-\frac{q^2}{4\,\kappa^2},\,0,\,\kappa^2\,z^2\right)}.
\end{eqnarray}

Here, for the process $H_1\to H_2\,\ell\,\bar{\nu}_\ell$, we chose $X=1$ and $X=2$ for the heavy-light mesons, and $\kappa$ denotes the weak-current energy scale. To evaluate this integral, we used the methods described in Appendix \ref{sec:Ap1}, yielding 

\begin{equation}\label{SW-form-factor}
    f_\pm\left(q^2\right)=\frac{8\,\kappa_1^2\,\kappa_2^2\,g_\pm}{\left(\kappa_1^2+\kappa^2_2\right)^2\,R^3\,\left(1-\frac{q^2}{4\,\kappa^2}\right)_2}\,_2F_1\left(-\frac{q^2}{4\,\kappa^2},\,2;\,3-\frac{q^2}{4\,\kappa^2};\,1-\frac{2\,\kappa^2}{\kappa^2_1+\kappa^2_2}\right),
\end{equation}

\noindent with $\left(x\right)_n$ the \emph{Pochhammer symbol}, and $_2F_1\left(a,\,b;\,c;\,x\right)$ the \emph{Gauss hypergeometric function} \cite{NIST:DLMF}. 

We fix the couplings $g_\pm$ using heavy quark symmetry, where the Isgur-Wise function is known, \emph{i.e.}, $\xi(q^2_{\text{max}})=1$. Thus, from Eq.~\eqref{eq:HQET-FF}, we obtain:

\begin{equation}
   f_\pm\left(q_\text{max}^2\right)=\pm\frac{\kappa_1\pm\kappa_2}{2\,\sqrt{\kappa_1\,\kappa_2}}, 
\end{equation}

\noindent with $q_\text{max}^2=6\,\left(\kappa_1-\kappa_2\right)^2$ in this particular case. This value should match Eq.~\eqref{SW-form-factor} when evaluated at $q_\text{max}^2$, allowing us to extract the value of $g_\pm$ as

\begin{equation}
    g_\pm=\pm\frac{\left(\kappa_1\pm\kappa_2\right)\left(\kappa^2_1+\kappa^2_2\right)^2\,\left(1+a_\text{max}\right)_2\,R^3}{16\,\left(\kappa_1\,\kappa_2\right)^{\frac{5}{2}}\,_2F_1\left(a_\text{max},\,2;\,3+a_\text{max};\,1-\frac{2\,\kappa^2}{\kappa^2_1+\kappa^2_2}\right)},
\end{equation}

\noindent where we have defined $a_\text{max}$ as

\begin{equation*}
    a_\text{max}=-\frac{3\,\left(\kappa_1-\kappa_2\right)^2}{2\,\kappa^2}.
\end{equation*}

The values of $g_\pm$ for $B\to D$ and $B_s\to D_s$ are summarized in Table~\ref{tab:FF}.

Collecting all of these results, we obtain a closed form for semileptonic form factors in the context of the Softwall Model as

\begin{equation}
    f^\text{SW}_\pm\left(q^2\right)=\pm\frac{\left(\kappa_1\pm\kappa_2\right)\,\left(1+a_\text{max}\right)_2}{2\,\left(\kappa_1\,\kappa_2\right)^{\frac{1}{2}}\,_2F_1\left(a_\text{max},\,2;\,3+a_\text{max};\,1-\frac{2\,\kappa^2}{\kappa^2_1+\kappa^2_2}\right)}\,\frac{_2F_1\left(-\frac{q^2}{4\,\kappa^2},\,2;\,3-\frac{q^2}{4\,\kappa^2};\,1-\frac{2\,\kappa^2}{\kappa^2_1+\kappa^2_2}\right)}{\,\left(1-\frac{q^2}{4\,\kappa^2}\right)_2}.
\end{equation}

Setting $q^2=0$, the semileptonic form factors reduce to a constant expressed in terms of $\kappa_1$, $\kappa_2$, and $\kappa$ alone. Therefore, after fixing $\kappa_1$ and $\kappa_2$ from the heavy-light meson spectroscopy, and given a well-known value of $f_\pm(0)$ (from experiment or lattice), we can determine the value of the mass scale $\kappa$ for the weak-current dual field:

\begin{equation}\label{kappa-fixing}
    f_\pm^\text{Exp., Lat.}(0)=\frac{\left(\kappa_1\pm\kappa_2\right)\,\left(1+a_\text{max}\right)_2}{2\,\left(\kappa_1\,\kappa_2\right)^{\frac{1}{2}}\,_2F_1\left(a_\text{max},\,2;\,3+a_\text{max};\,1-\frac{2\,\kappa^2}{\kappa^2_1+\kappa^2_2}\right)}.
\end{equation}

\noindent where the upper index $\text{Exp., Lat.}$ indicates that the left-hand side is obtained from experiments or lattice calculations. The bulk-to-boundary propagator has a pole at $q^2 = 4\kappa^2$. To avoid this pole in the physical range $q^2 \le q^2_{\text{max}} = 6(\kappa_1-\kappa_2)^2$, we require $4\kappa^2 > 6(\kappa_1-\kappa_2)^2$,  \emph{i.e.}, $\kappa > \frac{\sqrt{6}}{2}(\kappa_1-\kappa_2)$. From Tables~\ref{tab:SWM} and~\ref{tab:weak}, we find that this consistency check is satisfied since $\kappa=4.097\,\text{GeV}>1.705\,\text{GeV}$ for $B\to D$, and $\kappa=4.102\,\text{GeV}>1.699\,\text{GeV}$ for the $B_s\to D_s$ transition. 

Similarly, we can address $f_0(Q^2)$ following Eq.~\eqref{eqn:f0_holo}, yielding

\begin{eqnarray} \notag
    f_0^\text{SW}\left(q^2\right)&=&\left[g_++\frac{q^2}{6\,\left(\kappa_1^2-\kappa_2^2\right)}\,g_-\right]\,I\left(q^2\right)\\  \label{F_zero_SWM}
    &=&\frac{\left(\kappa_1+\kappa_2\right)\,\left(1+a_\text{max}\right)_2\,\left[1-\frac{q^2}{6\,\left(\kappa_1+\kappa_2\right)^2}\right]}{2\,\left(\kappa_1\,\kappa_2\right)^{\frac{1}{2}}\,_2F_1\left(a_\text{max},\,2;\,3+a_\text{max};\,1-\frac{2\,\kappa^2}{\kappa^2_1+\kappa^2_2}\right)}\,\times\\ \notag
    &\times&\frac{_2F_1\left(-\frac{q^2}{4\,\kappa^2},\,2;\,3-\frac{q^2}{4\,\kappa^2};\,1-\frac{2\,\kappa^2}{\kappa^2_1+\kappa^2_2}\right)}{\,\left(1-\frac{q^2}{4\,\kappa^2}\right)_2}.
\end{eqnarray}

Once we have extracted the form factor expressions, we can write the SWM Isgur-Wise function as 

\begin{equation}
    \xi_\text{SWM}\left(\omega\right)=\frac{\left(1+a_\text{max}\right)_2}{_2F_1\left(a_\text{max},\,2;\,3+a_\text{max};\,1-\frac{2\,\kappa^2}{\kappa^2_1+\kappa^2_2}\right)}\,\frac{_2F_1\left(-\frac{q(\omega)^2}{4\,\kappa^2},\,2;\,3+\frac{q(\omega)^2}{4\,\kappa^2};\,1-\frac{2\,\kappa^2}{\kappa^2_1+\kappa^2_2}\right)}{\,\left(1-\frac{q(\omega)^2}{4\,\kappa^2}\right)_2},
\end{equation}

\noindent where the transferred momentum is defined in terms of the recoil variable $\omega$ as follows

\begin{equation}
    q^2\left(\omega\right)=6\left(\kappa_1^2+\kappa^2_2\right)-12\,\kappa_1\,\kappa_2\,\omega.
\end{equation}

In Figure~\ref{fig:IWF}, we show the Softwall Model Isgur-Wise function for the $B\to D$ transition. Notice that near the maximum recoil point, \emph{i.e.}, $\omega_{\text{max}}=1.59$, the SWM prediction differs notably from the lattice result \cite{Bowler:2002zh}. Note that this point corresponds to $Q^2=0$, where subleading $1/m_Q$ corrections to the HQS result have not yet been included in our holographic computation.

A plot of the $f_+(q^2)$ and $f_0(q^2)$ form factors for the $B\to D$ and $B_s\to D_s$ transitions is shown in Figure~\ref{fig:FF-HQS}. Table~\ref{tab:FF} summarizes the main form-factor values, compared with experimental fits for $B\to D$ and lattice results for $B_s\to D_s$. The holographic approach, including HQS, shows deviations of around $4\,\%$ for the $ B\to D$ transition at $Q_{\text{max}}^2$ compared with the experimental fit \cite{Yao:2019vty}. A similar situation occurs in the $B_s\to D_s$ transition, where the deviations are less than $6\,\%$ compared with lattice results \cite{Monahan:2017uby}. However, compared with the pure HQS results derived from Eq.~\eqref{eq:HQET-FF}, the differences in both transitions are less than $2\,\%$. We attribute this discrepancy between the holographic predictions and the corresponding lattice/experimental data, as well as the deviation from pure HQS, to the absence of higher-order $1/m_Q$ corrections in our holographic computation.

Once we have the form factor expressions, we can straightforwardly calculate the decay rate for $H_1\to H_2\,\ell\,\bar{\nu}_\ell$ using Eq.~\eqref{eq:diff_decay_rate} for both light and heavy leptons. 

\begin{multline}
\frac{d \Gamma^\text{SW}}{d q^{2}}\left(H_1 \rightarrow H_2 \,l\, \bar{\nu}_{l}\right) = \frac{G_{F}^{2} |V_{CKM}|^{2}}{1152\,\sqrt{6}\, \pi^{3} \,\kappa_{1}^{3}} \left(1 - \frac{m_{\ell}^{2}}{q^{2}} \right) \times\\
\left[ \left(1 + \frac{m_{\ell}^{2}}{2\, q^{2}} \right) \left[\lambda^{\text{SW}}(q^{2})\right]^\frac{3}{2}\,\left|f^\text{SW}_{+}(q^{2})\right|^{2} + \frac{54\,m_{\ell}^{2}}{ q^{2}}\,\left(\kappa_{1}^{2} - \kappa_{2}^{2}\right)^{2}\,  \left[\lambda^{\text{SW}}(q^{2})\right]^\frac{1}{2}\,\left|f^\text{SW}_{0}(q^{2})\right|^{2} \right],
\end{multline}

\noindent where the kinematic region for this process is $m_\ell^2\leq q^2\leq6\,\left(\kappa_1-\kappa_2\right)^2$, and the Softwall Model version of the Källén function is given by: 

\begin{equation}
    \lambda^\text{SW}\left(q^2\right)=36\,\left(\kappa^2_1+\kappa^2_2\right)^2\left\{\left[1-\frac{q^2}{6\,\left(\kappa^2_1+\kappa^2_2\right)}\right]^2-\frac{4\,\kappa^2_1\,\kappa^2_2}{\left(\kappa^2_1+\kappa^2_2\right)^2}\right\}.
\end{equation}

Figure~\ref{fig:FF-HQS} shows the $B\to D$ and $B_s\to D_s$ differential decay rates for light leptons and the tau lepton.

This decay rate can be integrated in order to define the corresponding softwall \emph{branching fraction}

\begin{equation}
\label{BranchingRatio-SW}
\frac{\mathcal{B}^\text{SW}\left(H_1 \rightarrow H_2\, \ell \bar{\nu}_\ell\right)}{\tau\left(H_{1}\right)} = \int_{m_\ell^{2}}^{6\,\left(\kappa_{1} - \kappa_{2}\right)^{2}}\, {dq^{2}\, \frac{d \,\Gamma^\text{SW}}{d q^{2}}}\,\left(H_1 \rightarrow H_2 \,\ell\, \bar{\nu}_\ell\right),
\end{equation}

\noindent where $\tau\left(H_1\right)$ corresponds to the meson $H_1$ lifetime, which is an input in the model. For the $B^0$ and $B_s^0$ mesons, we used the PDG data: $\tau(B^0)=(1.5113\pm0.0027)\times10^{-12}$ s and $\tau(B_s^0)=(1.518\pm0.006)\times10^{-12}$ s respectively \cite{PDGroup:2026aaa}. Our findings for the branching fraction in each transition are 

\begin{eqnarray}
   \mathcal{B}^\text{SW}\left(B^0 \rightarrow D^+\, \mu^- \,\bar{\nu}_\mu\right) &=& 2.122\,\%,\\
    \mathcal{B}^\text{SW}\left(B_s^0 \rightarrow D_s^-\, \mu^+ \,\bar{\nu}_\mu\right) &=& 2.185\,\%.
\end{eqnarray}

These results are consistent with the experimental ones. For the $B\to D$ transition, the Heavy Flavor Averaging Group (HFLAV) \cite{HFLAV:2024ctg} reported $\mathcal{B}\left(B^0 \rightarrow D^+\, \mu^- \,\bar{\nu}_\mu\right)=2.12\pm0.02_\text{stat}\pm0.06_\text{syst}\,\%$, giving a relative error smaller than $1.0\,\%$. A similar situation occurs for $B_s\to D_s$. The LHCb Collaboration reported $\mathcal{B}\left(B_s^0 \rightarrow D_s^-\, \mu^+ \,\bar{\nu}_\mu\right)=2.49\pm0.12_\text{stat}\pm0.14_\text{syst}\pm0.16_\text{ext}\,\%$ \cite{LHCb:2020cyw}, implying a relative error smaller than $10\,\%$. 

Similarly, we calculate, for each transition, the lepton-flavor universality ratio

\begin{equation}
    R(D_{(s)})=\frac{\mathcal{B}\left(B_{(s)}\to D_{(s)}\,\tau\,\nu_\tau\right)}{\mathcal{B}\left(B_{(s)}\to D_{(s)}\,\ell\,\nu_\ell\right)},
\end{equation}

\noindent which measures the ratio of the branching fraction of the semileptonic decay to a $\tau$ lepton to the branching fraction to a light lepton $\ell$. The current experimental average is $R(D)_\text{exp}=0.342\pm0.026$ \cite{HFLAV:2024ctg} for $B\to D$, and the lattice result is $R(D_s)_\text{latt}=0.2989\pm0.0041$ \cite{Bordone:2025jur} for the $B_s\to D_s$ transition. At the time of writing this draft, there is no experimental report on $R(D_s)$. 

For $B\to D$, the SWM result for the ratio is $R_\text{SWM}(D)=0.299$, which is closer to the Standard Model prediction $R(D)=0.300\pm0.008$ \cite{Na:2015kha} than to the experimental average. This result is not surprising, since we impose HQS constraints to fix $g_\pm$ in the form factors. 

For the $B_s\to D_s$ transition, we obtain $R_\text{SWM}(D_s)=0.2965$, which has a relative error near $1.0\,\%$.

\subsection{Non-quadratic Softwall Model case}\label{sec:4.3}

\begin{figure}[htbp]
    \centering
    \includegraphics[width=0.49\linewidth]{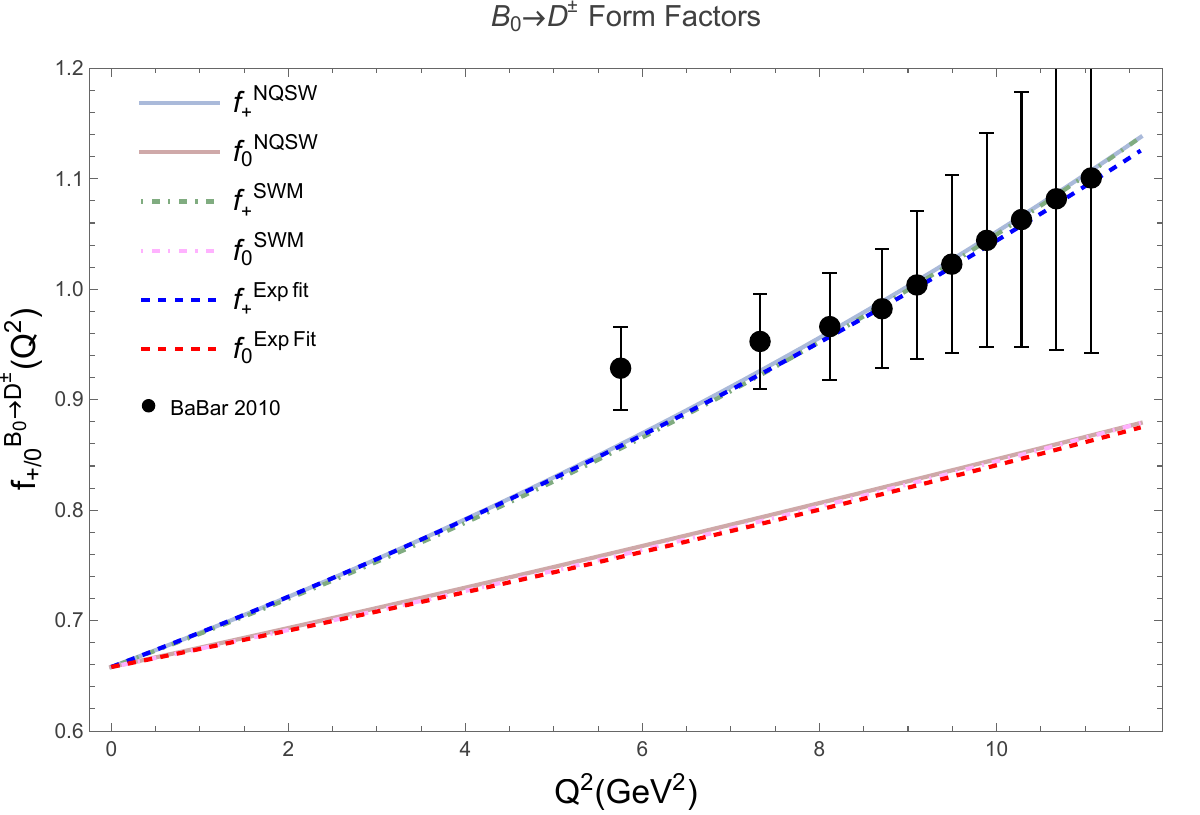}
    \includegraphics[width=0.49\linewidth]{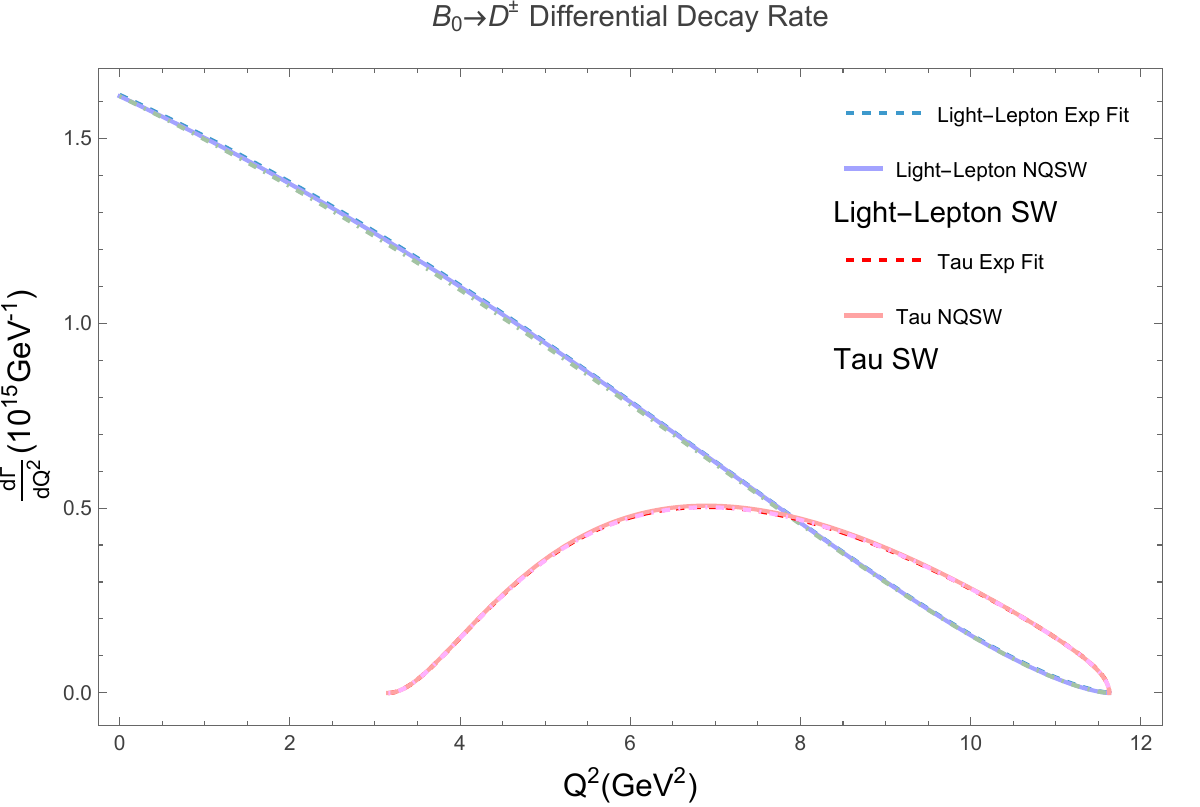}\\
    \includegraphics[width=0.49\linewidth]{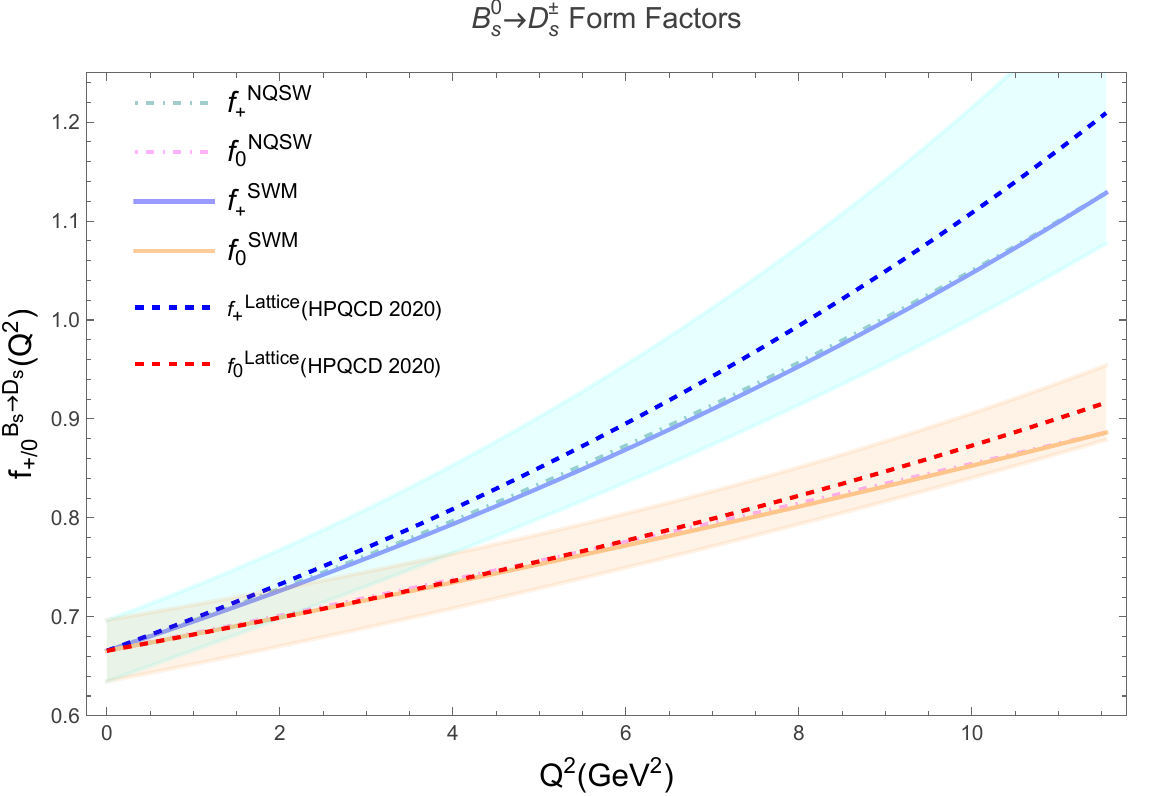}
    \includegraphics[width=0.49\linewidth]{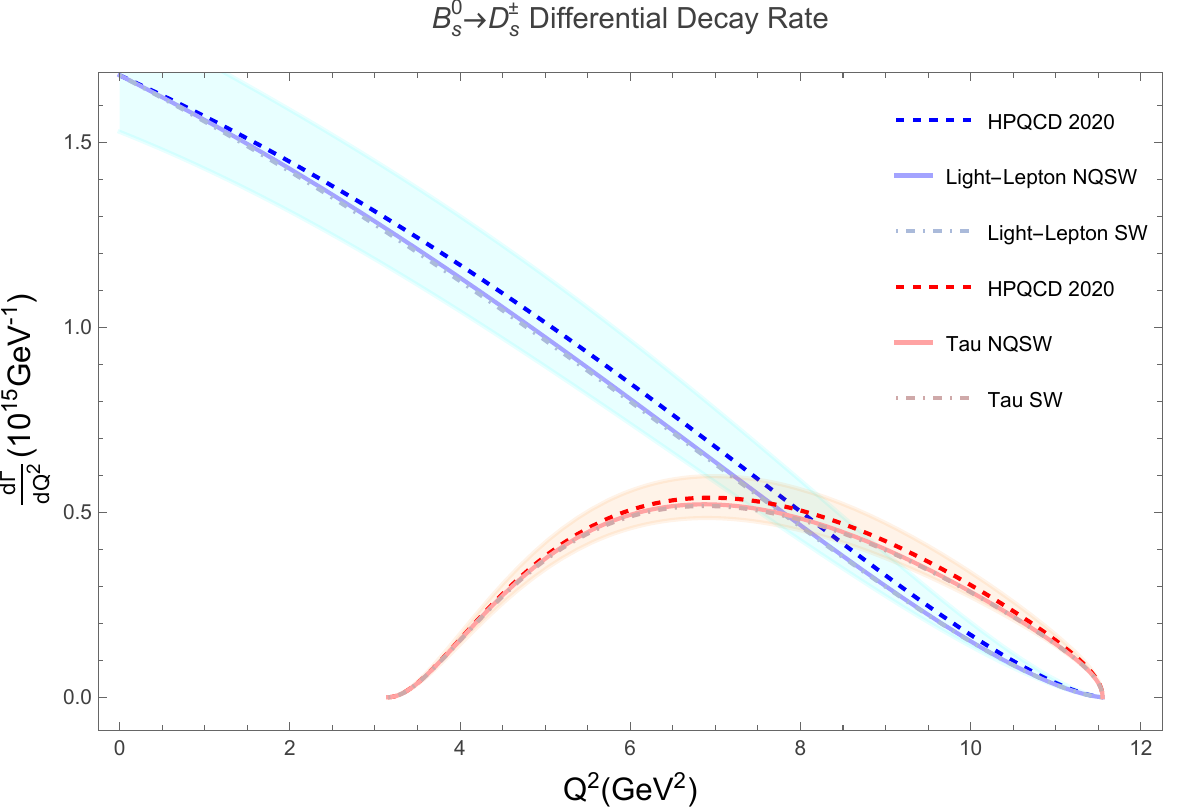}
    \caption{Form factors and differential decay rates for the $B\to D$ (top) and $B_s\to D_s$ (bottom) transitions. The solid lines correspond to the holographic predictions (SWM and NQSWM), the points are experimental data from BaBar \cite{BaBar:2010zpy}, the dashed lines are experimental fits for $B\to D$ \cite{Yao:2019vty}, and the shaded bands are lattice QCD results for $B_s\to D_s$ \cite{Monahan:2017uby, McLean:2019qcx}.}
    \label{fig:FF-HQS}
\end{figure}

The Softwall Model corresponds to linear Regge trajectories. This linear behavior characterizes light unflavored hadrons. However, when extended to the heavy quarkonium sector, the SWM does not adequately reproduce meson mass and electromagnetic decay-constant spectra \cite{Braga:2015jca}. In fact, in the heavy quarkonium sector, the trajectories can be either \emph{hydrogen-like} (as in non-relativistic QCD \cite{Lucha:1991vn}) or a general extension of the linear case, \emph{i.e.}, $M_n^2 = a (n+b)^\nu$, where $\nu$ measures the deviation from linearity. This behavior is also consistent with Bethe-Salpeter analyses of heavy mesons \cite{Chen:2018nnr, Chen:2018bbr, Chen:2018hnx}.

Holographically, the extension to non-linear radial Regge trajectories is achieved by the non-quadratic dilaton $\Phi(z) = (\kappa z)^{2-\alpha}$ \cite{MartinContreras:2020cyg}, where $\kappa$ fixes the Regge slope and $\alpha$ accounts for the effect of heavy quark mass: in the massless quark sector, $\alpha=0$ recovers the quadratic Softwall Model. Recall that, for heavy states, the Coulomb term in the Cornell potential is responsible for the short-distance dynamics and the fine structure. This translates into the non-linearity of the heavy-hadron spectrum, as shown by WKB analyses of the heavy-quark confining potential \cite{Quigg:1979vr}. Returning to the holographic framework, the Rydberg–Klein–Rees formula shows that $M_n^2 = a(n+b)^\nu$ corresponds to dilatons with the above structure \cite{MartinContreras:2025wnh}.

\begin{table}[htbp]
    \centering
    \footnotesize 
    \setlength{\tabcolsep}{4pt} 
    \begin{tabular}{c c c c c c c c c} 
        \multirow{2}{*}{\textbf{Quantity}} & \multicolumn{4}{c}{$B\to D$} & \multicolumn{4}{c}{$B_s\to D_s$} \\ 
        \cline{2-5} \cline{6-9} 
        & \textbf{NQSWM} & \textbf{SWM} & \textbf{Exp/Latt.} & \textbf{HQS} & \textbf{NQSWM} & \textbf{SWM} & \textbf{Exp/Latt.} & \textbf{HQS} \\
        \hline
        \hline
        $f_+(0)=f_0(0)$ & $0.658^*$ & $0.658^*$ & $0.658(17)$ \cite{Yao:2019vty} & N/A & $0.665^{**}$ & $0.665^{**}$ & $0.656(31)$ \cite{Monahan:2017uby} & N/A \\
        $f_+(q_\text{max}^2)$ & $1.138$ & $1.138$ & $1.190(71)$ \cite{Yao:2019vty} & $1.1377(4)$ & $1.128$ & $1.128$ & $1.205(45)$ \cite{Monahan:2017uby} & $1.1367(1)$ \\
        $f_0(q_\text{max}^2)$ & $0.879$ & $0.879$ & $0.900(34)$ \cite{Yao:2019vty} & $0.8789(3)$ & $0.886$ & $0.886$ & $0.913(44)$ \cite{Monahan:2017uby} & $0.879(3)$ \\
        $g_+$ & $0.691(1)$ & $1.661(1)$ & N/A & N/A & $0.681(1)$ & $1.743(1)$ & N/A & N/A \\
        $g_-$ & $-0.329(1)$ & $-0.793(1)$ & N/A & N/A & $-0.317(1)$ & $-0.832(1)$ & N/A & N/A
    \end{tabular}
    \caption{Form factors calculated in the Non-Quadratic (NQSWM) and Quadratic (SWM) Softwall models, compared with experimental or lattice fits for $B\to D$ \cite{Yao:2019vty} and $B_s\to D_s$ \cite{Monahan:2017uby} transitions, as well as Heavy Quark Symmetry (HQS) predictions at zero recoil summarized in Eq.~\eqref{eq:HQET-FF}. We compute the HQS values using the experimental masses quoted in Table~\ref{tab:NQSWM}. $^*$Input used to fix $\kappa$ for the $B\to D$ channel; $^{**}$Input used to fix $\kappa$ for the $B_s\to D_s$ channel.}
    \label{tab:FF}
\end{table}

Once the pairs $(\kappa_X, \alpha_X)$ are fixed for both hadrons and the weak-current parameter $\kappa$ is determined (as explained in Sec.~\ref{sec.4.1}), we use the Nelder–Mead method to compute the ground-state masses for the $B^0$, $D^\pm$, $B_s^0$, and $D_s^\pm$ mesons. These masses fix the kinematics at the zero-recoil point, where we fix $g_\pm$ according to Eq.~\eqref{eq:constraint}. We then compute the integral $I(q^2)$ numerically.

\begin{figure}
    \centering
    \includegraphics[width=0.65\linewidth]{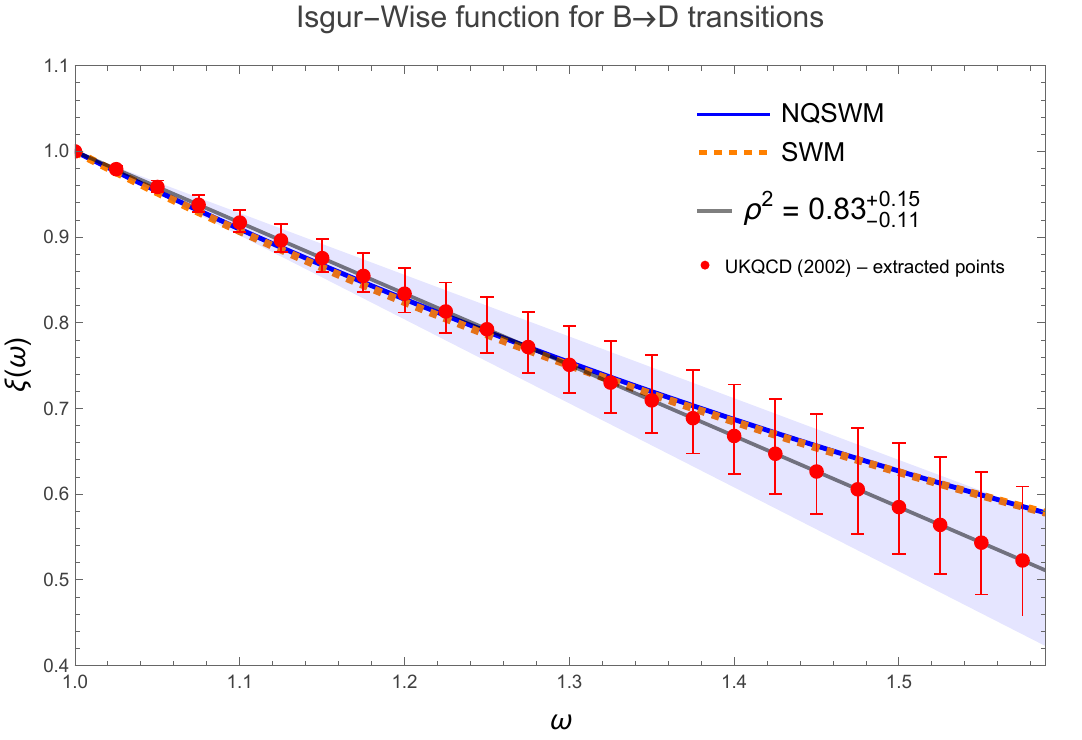}
    \caption{The Isgur-Wise function $\xi(\omega)$ from the UKQCD Collaboration. The gray solid line is the result of a simultaneous fit to the form factors $h_+(\omega)$ (for $B\to D$) and $h_{A_a}(\omega)$ (for $B\to D^*$), assuming the linear parametrization $\xi(\omega)=1-\rho^2(\omega-1)$ \cite{Bowler:2002zh}, where the weak vector current is written as $\langle D\left|J^\mu\right|B\rangle\propto h_+(\omega)\,(v_1+v_2)\mu+h_-(\omega)\,(v_1-v_2)^\mu$. The shaded bands indicate the $1\,\sigma$ uncertainty from the fit, giving $\rho=0.83^{+0.15}_{-0.11}$. The blue and orange continuous lines correspond to the Non-Quadratic (NQSWM) and the Quadratic Softwall Model (SWM).}
    \label{fig:IWF}
\end{figure}

The kinematic region is defined by $m_\ell^2 \leq q^2 \leq q_{\text{max}}^2$, where the upper limit is evaluated using the masses obtained from the NQSWM, \emph{i.e.}, $q_{\text{max}}^2 \equiv (\tilde{M}_B - \tilde{M}_D)^2$, where $\tilde{M}_X$ are the mass outcomes from the Schrödinger-like equation \eqref{schr-equn} in the NQSWM scheme. For reference, in the $B\to D$ transition, the experimental value is $q^2_{\text{max, exp}} = 11.629 \pm 2.106\ \text{GeV}^2$, while the corresponding theoretical prediction from the NQSWM is $q^2_{\text{max, th}} = 11.628\ \text{GeV}^2$.

Figure~\ref{fig:FF-HQS} shows the scalar and vector form factors for the $B\to D$ transition. We also plot a parametrization for $f_+$ and $f_0$ that combines experimental and lattice data \cite{Yao:2019vty}, along with the experimental data from BaBar \cite{BaBar:2010zpy}. In Table~\ref{tab:FF}, we summarize the characteristic values at zero and maximum recoil, compared with the experimental fit and the HQS theoretical limit. As mentioned in Sec.~\ref{sec.4.1}, the condition $f_0(0) = f_+(0) = 0.658 \pm 0.017$ is used to fix the numerical value of $\kappa$. At the zero-recoil point, the values of $f_{+,0}(q_{\text{max}}^2)$ have a relative error smaller than $4\,\%$. We perform a similar analysis for the $B_s\to D_s$ transition, where the form factor values at $q_{\text{max}}^2$ have a relative error of around $6\,\%$. In both cases, the deviations arise from omitting heavy-quark symmetry corrections.

Similarly, we compute the Isgur–Wise function for the $B\to D$ transition following Eq.~\eqref{eq:IWF}. Figure~\ref{fig:IWF} shows our results, where, as in the SWM case, we observe noticeable differences near $\omega_{\text{max}}$. In both holographic models, we include no corrections beyond HQS.

The discrepancy between the holographic IWF and the lattice QCD result at maximum recoil reflects the sensitivity of the overlap integral $I(q^2)$ to the detailed shape of the confining potential. While the dilaton parameters are fixed by the meson mass spectrum, they do not uniquely determine the wavefunction at large $z$ (small $q^2$). The absence of higher-order $1/m_Q$ corrections in the holographic action—such as derivative couplings or subleading terms in the bulk mass—prevents the model from fully capturing the $q^2$-dependence of the form factors at large recoil.

As in the SWM case, we also computed the differential decay rate for the light-lepton and tau-lepton cases for the $B\to D$ and $B_s\to D_s$ transitions. Figure~\ref{fig:FF-HQS} shows the results. In this case, the kinematic region is $m_\ell^2 \leq q^2 \leq 11.628\,\text{GeV}^2$.

Integrating this differential decay rate, we extract the branching fractions in the NQSWM approach:

\begin{eqnarray}
    \mathcal{B}^\text{NQSWM}\left(B^0 \rightarrow D^- \, \mu^+ \, \bar{\nu}_\mu\right) &=& 2.135\,\%,\\
    \mathcal{B}^\text{NQSWM}\left(B_s^0 \rightarrow D_s^- \, \mu^+ \, \bar{\nu}_\mu\right) &=& 2.607\,\%.
\end{eqnarray}

Furthermore, the ratio of the exclusive $B_s^0\to D_s^- \, \mu^+ \, \bar{\nu}_\mu$ branching fraction to its $B^0$ counterpart is given by

\begin{equation}
    \mathcal{R}=\frac{\mathcal{B}^\text{NQSWM}\left(B_s^0 \rightarrow D_s^- \, \mu^+ \, \bar{\nu}_\mu\right)}{\mathcal{B}^\text{NQSWM}\left(B^0 \rightarrow D^- \, \mu^+ \, \bar{\nu}_\mu\right)}=1.221,
\end{equation}

\noindent which is consistent with the reported value in the literature $\mathcal{R}=1.09\pm0.05_\text{stat}\pm0.06_\text{syst}\pm0.05_\text{ext}$ \cite{Yadav:2024zrn}. Similarly, for the lepton-flavor universality ratio defined in Eq.~\eqref{eq:R_definition}, we obtain $R_\text{NQSWM}(D)=0.299$ and $R_\text{NQSWM}(D_s)=0.327$, in agreement with the experimental, lattice, and SM predictions.

\subsection{Testing the universality of the Isgur-Wise function}

\begin{figure}[htbp]
    \centering
    \includegraphics[width=0.49\linewidth]{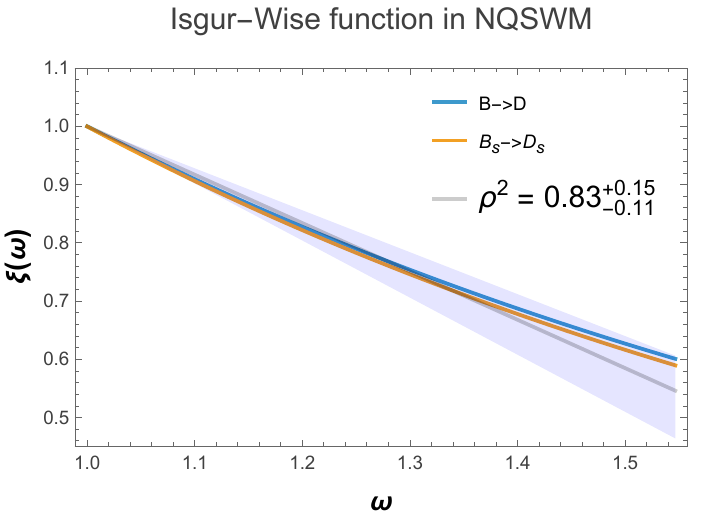}
    \includegraphics[width=0.49\linewidth]{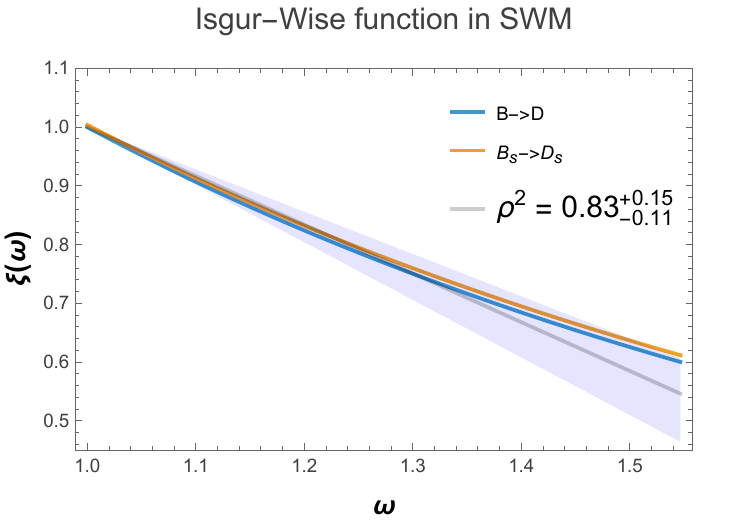}
    \caption{Isgur--Wise function (IWF) for $b \to c$ transitions in the NQSWM (left) and SWM (right). The blue and orange curves show the IWF for $B \to D$ and $B_s \to D_s$, respectively, with HQS imposed at zero recoil. The gray curve and the light purple band correspond to the lattice QCD result from Ref.~\cite{Monahan:2017uby}. The light purple band represents the $1\,\sigma$ uncertainty of the lattice determination. The near-overlap of the HQS-constrained curves demonstrates the universality of the IWF.}
    \label{fig:IWF-Universality}
\end{figure}

An interesting test is the \emph{universality of the Isgur--Wise function} in $b \to c$ transitions. In the limit where the heavy-quark masses $ m_b$ and $ m_c$ go to infinity, QCD exhibits an approximate spin-flavor symmetry, the so-called Heavy Quark Symmetry (HQS). The emergence of HQS implies that the strong-interaction dynamics of the light degrees of freedom, \emph{i.e.}, light quarks and gluon clouds, are completely independent of the flavor and spin of the heavy quarks. A direct consequence of HQS is the Isgur--Wise function itself, which is a \emph{single} and \emph{universal} function of the velocity transfer $\omega = v_1 \cdot v_2$ that encodes all the non-perturbative dynamics. Thus, the same Isgur--Wise function appears in the form factors for any transition between heavy-light mesons, where the light quarks act as mere spectators. Therefore, for transitions such as $B \to D$, $B_s \to D_s$, $B \to D^*$, or $\Lambda_b \to \Lambda_c$, HQS predicts the same IWF. This is the so-called \emph{universality of the Isgur--Wise function} \cite{Colangelo:1998fi, Chow:1996xf}. Figure~\ref{fig:IWF-Universality} illustrates the holographic version of this concept.

In the strict heavy-quark limit, the Isgur--Wise function extracted from the $B \to D$ and $B_s \to D_s$ transitions should coincide, reflecting the universality predicted by HQET. However, in our holographic models, we use finite heavy quark masses, which introduce subleading $1/m_Q$ corrections. When we impose HQS constraints at zero recoil, we effectively enforce this universality, and the two channels yield the same $\xi(\omega)$, up to small deviations from different dilaton parameters for the $B$ and $D$ mesons.

At the maximum recoil point, $\omega_{\text{max}} \approx 1.55$ for the $B_s \to D_s$ transition, we observe a discrepancy between the holographic IWF, $\xi^{\text{hol}}(\omega_{\text{max}}) = 0.578$, and the lattice QCD determination, $\xi^{\text{latt}}(\omega_{\text{max}}) = 0.511$ \cite{Monahan:2017uby}. This deviation reflects the limitations of our simplified holographic model at large recoil, where the light-cloud overlap is most sensitive to the confining potential and heavy-quark dynamics. Near zero recoil, where HQS is most reliable, the model agrees much better with lattice data. This suggests that the holographic framework captures the leading dynamics but requires subleading $1/m_Q$ corrections to fully reproduce the $q^2$-dependence of the form factors at large recoil.

This picture indicates that the IWF shape is largely determined by the holographic dynamics encoded in the overlap integral $I(q^2)$, which is fixed by the meson spectroscopy. The independent fit of $g_\pm$ primarily adjusts the overall normalization and the relative weight of the $f_-$ component, but does not significantly alter the intrinsic $q^2$-dependence predicted by the model. This stability at $\xi(\omega_{\text{max}})$ suggests that the holographic model robustly describes the form-factor shape, and the remaining discrepancy with lattice QCD stems from the absence of higher-order $1/m_Q$ corrections in our calculation.

\section{Conclusions}\label{sec:Concl}
In this work, we have computed the Isgur--Wise function and the associated semileptonic form factors for the $B\to D$ and $B_s\to D_s$ transitions using two bottom-up holographic QCD models: the Softwall Model (SWM) with a quadratic dilaton and the Non-Quadratic Softwall Model (NQSWM) with a generalized dilaton profile. We fixed the holographic model parameters using the meson mass spectra, and extracted the form factors \emph{via} overlap integrals of bulk wavefunctions. We imposed Heavy Quark Symmetry (HQS) at zero recoil.

Our main findings are as follows. First, the HQS-constrained models reproduce the universality of the Isgur--Wise function, with the $B \to D$ and $B_s \to D_s$ transitions yielding nearly identical $\xi(\omega)$ within the holographic framework. This shows that the holographic approach, when constrained by HQS, is consistent with Heavy Quark Effective Theory predictions. 

Second, we have addressed the apparent paradox of a non-conserved electroweak current dual to a massless bulk vector field. We showed that the non-conservation of the electroweak current arises from explicit flavor-symmetry breaking by quark masses and does not generate an anomalous dimension. Consequently, the conformal dimension of the current remains exactly $\Delta_J = 3$, and the holographic dictionary assigns it a massless bulk field. The non-conservation is encoded not in a mass term for the bulk field, but in the independence of the boundary couplings $g_+$ and $g_-$, which reflects the modified Ward identity. This is the holographic realization of the statement that \emph{the non-conservation is an infrared effect that does not modify the ultraviolet properties of the operator}.

Third, our results for the lepton-flavor universality ratios $R(D)$ and $R(D_s)$ are consistent with the Standard Model and lattice QCD. The branching fractions for $B_s \to D_s$ are in good agreement with LHCb data within the expected uncertainties of the model.

Finally, we observe that the largest discrepancies between the holographic predictions and lattice QCD data occur near the maximum recoil point ($\omega_{\text{max}} \approx 1.55$). This suggests that the holographic models capture the leading dynamics of the Isgur--Wise function but require additional corrections to fully reproduce the $q^2$--dependence of the form factors at large recoil. In the holographic framework, these corrections correspond to a small bulk mass $M_5^2 \sim \mathcal{O}(1/m_Q)$, generated by mixing the leading HQET current with subleading operators. At the order considered here, such terms are neglected, and the massless bulk-to-boundary propagator remains the appropriate description.

Future work should focus on extending the analysis to other heavy-flavor transitions, such as $B \to D^*$ and $\Lambda_b \to \Lambda_c$, to further test the universality of the Isgur--Wise function.

\begin{acknowledgments}
M. A. Martin Contreras would like to acknowledge the financial support provided by the National Natural Science Foundation of China (NSFC) under grant No. 12350410371. A. Vega is partially supported by the Centro de Física Teórica de Valparaíso (CeFiTeV) and by FONDECYT (Chile) under grant No. 1251106.
\end{acknowledgments}

\appendix

\section{Gauss Hypergeometric functions} \label{sec:Ap1}
In the context of form factors in the KKSS model, this type of integral may arise:

\begin{widetext}
    \begin{equation}
    I_n\left(q^2\right)=\int_0^\infty{dz\,z^n\,e^{-\lambda^2\,z^2}\,\Gamma\left(1-\frac{q^2}{4\,\kappa^2}\right)\,U\left(-\frac{q^2}{4\,\kappa^2},\,0,\,\kappa^2\,z^2\right)},\,\,\text{with }n\in\mathbb{Z}^+.
\end{equation}
\end{widetext}

It is customary to consider the variable change $x=\kappa^2\,z^2$, $p=\lambda^2/\kappa^2$, and $a=-q^2/4\,\kappa^2$ such that 

\begin{equation}
    I_n\left(a\right)=\frac{\Gamma\left(1+a\right)}{2\,\kappa^{n+1}}\,\int_0^\infty{dx\,x^{\frac{n-1}{2}}\,e^{-p\,x}}\,U\left(a,\,0,\,x\right).
\end{equation}

To evaluate this Integral involving the Tricomi Function $U(a,\,b,\,x)$, we consider first its integral representation 

\begin{equation}
    U\left(a,\,b,\,x\right)=\frac{1}{\Gamma\left(a\right)}\int_0^\infty{dt\,e^{-x\,t}\,t^{a-1}\,\left(1+t\right)^{-a+b-1}}.
\end{equation}

\noindent fixing $b=0$, we can write $I_n(a)$ as follows

\begin{eqnarray*}
   I_n(a)&=& \frac{\Gamma\left(1+a\right)}{2\,\kappa^{n+1}\,\Gamma\left(a\right)}\,\int_0^\infty{dx\,x^{\frac{n-1}{2}}\,e^{-p\,x}\,\int_0^\infty{dt\,e^{-x\,t}\,t^{a-1}\,\left(1+t\right)^{-a-1}}}\\
   &=&\frac{\Gamma\left(1+a\right)}{2\,\kappa^{n+1}\,\Gamma\left(a\right)}\,\int_0^\infty{dt\,t^{a-1}\,\left(1+t\right)^{-a-1}\,\int_0^\infty{dx\,x^{\frac{n-1}{2}}\,e^{-\left(p+t\right)\,x}}}\\
   &=&\frac{\Gamma\left(1+a\right)\,\Gamma\left(\frac{n+1}{2}\right)}{2\,\kappa^{n+1}\,\Gamma\left(a\right)}\,\int_0^\infty{dt\,t^{a-1}\,\frac{\left(1+t\right)^{-a-1}}{\left(p+t\right)^{\frac{n+1}{2}}}}
\end{eqnarray*}

The \emph{trick} to solve the integral is to transform the integration limit from $[0,\,\infty)$ into $[0,\,1]$ by using the change

\begin{equation*}
    t=\frac{u}{1-u}
\end{equation*}

\noindent and rewriting the $I_n(a)$ integral as: 

\begin{equation}
    I_n(a)=\frac{\Gamma\left(1+a\right)\,\Gamma\left(\frac{n+1}{2}\right)}{2\,\kappa^{n+1}\,\Gamma\left(a\right)} \,\int_0^1{du\,\frac{u^{a-1}\,\left(1-u\right)^{\frac{n+1}{2}}}{\left[p-(p-1)\,u\right]^{\frac{n+1}{2}}}}
\end{equation}

The integral in the last expression is the integral representation of the \emph{Gauss hypergeometric function}  $_2F_1(\alpha,\,\beta;\,\gamma,\,
z)$ \cite[\href{https://dlmf.nist.gov/15.2}{§15.6.1}]{NIST:DLMF}: 

\begin{equation}
    _2F_1\left(\alpha,\,\beta;\,\gamma;\,z\right)=\frac{\Gamma\left(\gamma\right)}{\Gamma\left(\beta\right)\,\Gamma\left(\gamma-\beta\right)}\,\int_0^1{du\,u^{\beta-1}\,\left(1-u\right)^{\gamma-\beta-1}\,\left(1-u\,z\right)^{-\alpha}}
\end{equation}

If we now fix the hypergeometric indices  as:

\begin{equation*}
    \alpha=\frac{n+1}{2},\,\,\beta=a,\,\,\gamma=\frac{n+3}{2}+a,\,\,z=1-\frac{1}{p}
\end{equation*}

\noindent the integral $I_n(a)$ acquires the form

\begin{equation}
    I_n\left(a\right)=\frac{\Gamma\left(\frac{n+3}{2}\right)\,\Gamma\left(\frac{n+1}{2}\right)}{2\,\left(1+a\right)_\frac{n+1}{2}\,\lambda^{n+1}}\,_2F_1\left(a,\,\frac{n+1}{2};\,\frac{n+3}{2}+a;\,1-\frac{\kappa^2}{\lambda^2}\right)
\end{equation}

\noindent where we have used the $_2F_1$ symmetry properties, and $(x)_\nu$ defines the Pochhammer symbols. This form simplifies for specific values of $n$. For example, in the $n=3$ case, we obtain:

\begin{equation}
    I_3\left(a\right)=\frac{1}{(1+a)\left(2+a\right)\,\lambda^4}\,_2F_1\left(a,\,2;\,3+a;\,1-\frac{\kappa^2}{\lambda^2}\right)
\end{equation}

\noindent which is a closed form. The \emph{Pfaff theorem} \cite[\href{https://dlmf.nist.gov/15.2}{§15.8.1}]{NIST:DLMF} can be used to simplify the hypergeometric function for specific values of $a$. 
\bibliographystyle{unsrt}
\bibliography{references}

@PREAMBLE{
 "\providecommand{\noopsort}[1]{}" 
 # "\providecommand{\singleletter}[1]{#1}%" 
}

@book{KimPham,
      author         = "Quang Ho-Kim and Xuan-Yem Pham",
       title          = "",
      journal        = "Elementary Particles and their Interactions: Concepts and Phenomena, Springer",
      volume         = "",
      year           = "1998",
      pages          = "",
      doi            =  "10.1007/978-3-662-03712-6",
      eprint         = "",
      archivePrefix  = "",
      primaryClass   = "",
      reportNumber   = "",
      SLACcitation   = "%%CITATION = ARXIV:1310.8555;%%",
      publisher = "Springer Nature"
}

@book{Narison,
      author         = "Stephan Narison",
       title          = "",
      journal        = "QCD as a Theory of Hadrons: From Partons to Confinement, Cambridge Monographs on Particle Physics, Nuclear Physics and Cosmology",
      volume         = "17",
      year           = "2004",
      pages          = "",
      doi            =  "",
      eprint         = "",
      archivePrefix  = "",
      primaryClass   = "",
      reportNumber   = "ISBN 978-0-521-03731-0",
      SLACcitation   = "%%CITATION = ARXIV:1310.8555;%%",
      publisher="Cambridge University Press"
}

@article{Burdman:1993es,
    author = "Burdman, Gustavo and Ligeti, Zoltan and Neubert, Matthias and Nir, Yosef",
    title = "{The Decay B ---\ensuremath{>} pi lepton neutrino in heavy quark effective theory}",
    eprint = "hep-ph/9309272",
    archivePrefix = "arXiv",
    reportNumber = "SLAC-PUB-6345, WIS-93-86-PH, UMHEP-382A",
    doi = "10.1103/PhysRevD.49.2331",
    journal = "Phys. Rev. D",
    volume = "49",
    pages = "2331--2345",
    year = "1994"
}

@article{Isgur:1990kf,
    author = "Isgur, Nathan and Wise, Mark B.",
    title = "{Relationship Between Form-factors in Semileptonic $\bar{B}$ and $D$ Decays and Exclusive Rare $\bar{B}$ Meson Decays}",
    reportNumber = "CALT-68-1625, UTPT-90-02",
    doi = "10.1103/PhysRevD.42.2388",
    journal = "Phys. Rev. D",
    volume = "42",
    pages = "2388--2391",
    year = "1990"
}

@article{MartinContreras:2020cyg,
    author = "Martin Contreras, Miguel Angel and Vega, Alfredo",
    title = "{Nonlinear Regge trajectories with AdS/QCD}",
    eprint = "2004.10286",
    archivePrefix = "arXiv",
    primaryClass = "hep-ph",
    doi = "10.1103/PhysRevD.102.046007",
    journal = "Phys. Rev. D",
    volume = "102",
    number = "4",
    pages = "046007",
    year = "2020"
}

@article{Isgur:1989vq,
    author = "Isgur, Nathan and Wise, Mark B.",
    title = "{Weak Decays of Heavy Mesons in the Static Quark Approximation}",
    reportNumber = "UTPT-89-27, CALT-68-1585",
    doi = "10.1016/0370-2693(89)90566-2",
    journal = "Phys. Lett. B",
    volume = "232",
    pages = "113--117",
    year = "1989"
}

@inproceedings{Wise:1997sg,
    author = "Wise, Mark B.",
    title = "{Heavy quark physics: Course}",
    booktitle = "{Les Houches Summer School in Theoretical Physics, Session 68: Probing the Standard Model of Particle Interactions}",
    eprint = "hep-ph/9805468",
    archivePrefix = "arXiv",
    reportNumber = "CALT-68-2172",
    pages = "1051--1089",
    month = "7",
    year = "1997"
}

@misc{NIST:DLMF,
  key = "{\relax DLMF}",
  title = "{\it NIST Digital Library of Mathematical Functions}",
  howpublished = "\url{https://dlmf.nist.gov/}, Release 1.2.7 of 2026-06-15",
  url = "https://dlmf.nist.gov/",
  note = "F.~W.~J. Olver, A.~B. {Olde Daalhuis}, D.~W. Lozier, B.~I. Schneider, R.~F. Boisvert, C.~W. Clark, B.~R. Miller, B.~V. Saunders, H.~S. Cohl, and M.~A. McClain, eds."
}

@article{Karch:2006pv,
    author = "Karch, Andreas and Katz, Emanuel and Son, Dam T. and Stephanov, Mikhail A.",
    title = "{Linear confinement and AdS/QCD}",
    eprint = "hep-ph/0602229",
    archivePrefix = "arXiv",
    reportNumber = "BUHEP-06-02, INT-PUB-06-04",
    doi = "10.1103/PhysRevD.74.015005",
    journal = "Phys. Rev. D",
    volume = "74",
    pages = "015005",
    year = "2006"
}

@article{Batell:2008zm,
    author = "Batell, Brian and Gherghetta, Tony",
    title = "{Dynamical Soft-Wall AdS/QCD}",
    eprint = "0801.4383",
    archivePrefix = "arXiv",
    primaryClass = "hep-ph",
    reportNumber = "UMN-TH-2535-08",
    doi = "10.1103/PhysRevD.78.026002",
    journal = "Phys. Rev. D",
    volume = "78",
    pages = "026002",
    year = "2008"
}

@article{Li:2013oda,
    author = "Li, Danning and Huang, Mei",
    title = "{Dynamical holographic QCD model for glueball and light meson spectra}",
    eprint = "1303.6929",
    archivePrefix = "arXiv",
    primaryClass = "hep-ph",
    doi = "10.1007/JHEP11(2013)088",
    journal = "JHEP",
    volume = "11",
    pages = "088",
    year = "2013"
}

@article{Chen:2024ckb,
    author = "Chen, Xun and Huang, Mei",
    title = "{Machine learning holographic black hole from lattice QCD equation of state}",
    eprint = "2401.06417",
    archivePrefix = "arXiv",
    primaryClass = "hep-ph",
    doi = "10.1103/PhysRevD.109.L051902",
    journal = "Phys. Rev. D",
    volume = "109",
    number = "5",
    pages = "L051902",
    year = "2024"
}

@article{LHCb:2020cyw,
    author = "Aaij, Roel and others",
    collaboration = "LHCb",
    title = "{Measurement of $|V_{cb}|$ with $B_s^0 \to D_s^{(*)-} \mu^+ \nu_{\mu}$ decays}",
    eprint = "2001.03225",
    archivePrefix = "arXiv",
    primaryClass = "hep-ex",
    reportNumber = "LHCb-PAPER-2019-041, CERN-EP-2019-282",
    doi = "10.1103/PhysRevD.101.072004",
    journal = "Phys. Rev. D",
    volume = "101",
    number = "7",
    pages = "072004",
    year = "2020"
}

@article{HFLAV:2024ctg,
    author = "Banerjee, Sw. and others",
    collaboration = "Heavy Flavor Averaging Group (HFLAV)",
    title = "{Averages of b-hadron, c-hadron, and {\ensuremath{\tau}}-lepton properties as of 2023}",
    eprint = "2411.18639",
    archivePrefix = "arXiv",
    primaryClass = "hep-ex",
    doi = "10.1103/x87q-tld5",
    journal = "Phys. Rev. D",
    volume = "113",
    number = "1",
    pages = "012008",
    year = "2026"
}

@article{McLean:2019qcx,
    author = "McLean, E. and Davies, C. T. H. and Koponen, J. and Lytle, A. T.",
    title = "{$B_s\to D_s \ell\nu$ Form Factors for the full $q^2$ range from Lattice QCD with non-perturbatively normalized currents}",
    eprint = "1906.00701",
    archivePrefix = "arXiv",
    primaryClass = "hep-lat",
    doi = "10.1103/PhysRevD.101.074513",
    journal = "Phys. Rev. D",
    volume = "101",
    number = "7",
    pages = "074513",
    year = "2020"
}

@article{Braga:2015jca,
    author = "Braga, Nelson R. F. and Martin Contreras, M. A. and Diles, Saulo",
    title = "{Decay constants in soft wall AdS/QCD revisited}",
    eprint = "1507.04708",
    archivePrefix = "arXiv",
    primaryClass = "hep-th",
    doi = "10.1016/j.physletb.2016.10.046",
    journal = "Phys. Lett. B",
    volume = "763",
    pages = "203--207",
    year = "2016"
}

@article{Lucha:1991vn,
    author = "Lucha, W. and Schoberl, F. F. and Gromes, D.",
    title = "{Bound states of quarks}",
    doi = "10.1016/0370-1573(91)90001-3",
    journal = "Phys. Rept.",
    volume = "200",
    pages = "127--240",
    year = "1991"
}

@article{Chen:2018nnr,
    author = "Chen, Jiao-Kai",
    title = "{Regge trajectories for the mesons consisting of different quarks}",
    doi = "10.1140/epjc/s10052-018-6134-0",
    journal = "Eur. Phys. J. C",
    volume = "78",
    number = "8",
    pages = "648",
    year = "2018"
}

@article{Chen:2018bbr,
    author = "Chen, Jiao-Kai",
    title = "{Concavity of the meson Regge trajectories}",
    eprint = "1807.11003",
    archivePrefix = "arXiv",
    primaryClass = "hep-ph",
    doi = "10.1016/j.physletb.2018.10.022",
    journal = "Phys. Lett. B",
    volume = "786",
    pages = "477--484",
    year = "2018"
}

@article{Chen:2018hnx,
    author = "Chen, Jiao-Kai",
    title = "{Regge trajectories for heavy quarkonia from the quadratic form of the spinless Salpeter-type equation}",
    doi = "10.1140/epjc/s10052-018-5718-z",
    journal = "Eur. Phys. J. C",
    volume = "78",
    number = "3",
    pages = "235",
    year = "2018"
}

@article{Quigg:1979vr,
    author = "Quigg, C. and Rosner, Jonathan L.",
    title = "{Quantum Mechanics with Applications to Quarkonium}",
    reportNumber = "FERMILAB-PUB-79-022-T",
    doi = "10.1016/0370-1573(79)90095-4",
    journal = "Phys. Rept.",
    volume = "56",
    pages = "167--235",
    year = "1979"
}

@unpublished{MartinContreras:2025wnh,
    author = "Martin Contreras, Miguel Angel and Vega, Alfredo",
    title = "{Engineering Confining Dilatons: A WKB Inverse Problem in Holographic QCD}",
    eprint = "2509.04956",
    archivePrefix = "arXiv",
    primaryClass = "hep-ph",
    month = "9",
    year = "2025"
}

@article{GaoHan2012,
    author = {Gao, F. and Han, L.},
    title = {Implementing the Nelder-Mead simplex algorithm with adaptive parameters},
    journal = {Computational Optimization and Applications},
    volume = {51},
    number = {1},
    pages = {259-277},
    year = {2012},
    doi = {10.1007/s10589-010-9329-3}
}

@article{PDGroup:2026aaa,
    author = "Takahashi, F. and others",
    collaboration = "Particle Data Group",
    title = "{Review of Particle Physics}",
    doi = "10.1142/S0217751X26300115",
    journal = "Int. J. Mod. Phys. A",
    volume = "41",
    pages = "2630011",
    year = "2026"
}

@article{Gupta:2018ydm,
    author = "Gupta, Pallavi and Upadhyay, A.",
    title = "{Masses and strong decay properties of radially excited bottom states B(2S) and B(2P) with their strange partners B$_{s}$(2S) and B$_{s}$(2P)}",
    eprint = "1806.05366",
    archivePrefix = "arXiv",
    primaryClass = "hep-ph",
    doi = "10.1140/epja/i2018-12590-6",
    journal = "Eur. Phys. J. A",
    volume = "54",
    number = "9",
    pages = "160",
    year = "2018"
}

@article{LHCb:2015aaf,
    author = "Aaij, Roel and others",
    collaboration = "LHCb",
    title = "{Precise measurements of the properties of the $B_1(5721)^{0,+}$ and $B^\ast_2(5747)^{0,+}$ states and observation of $B^{+,0}\pi^{-,+}$ mass structures}",
    eprint = "1502.02638",
    archivePrefix = "arXiv",
    primaryClass = "hep-ex",
    reportNumber = "LHCB-PAPER-2014-067, CERN-PH-EP-2015-021",
    doi = "10.1007/JHEP04(2015)024",
    journal = "JHEP",
    volume = "04",
    pages = "024",
    year = "2015"
}

@article{refId0,
	author = {{Hoque, Rashidul} and {Hazarika, B. J.} and {Choudhury, D. K.}},
	title = {2S and 3S State Masses and decay constants of heavy-flavour mesons in a non-relativistic QCD potential model with three-loop effects in V-scheme},
	DOI= "10.1140/epjc/s10052-020-08756-4",
	url= "https://doi.org/10.1140/epjc/s10052-020-08756-4",
	journal = {Eur. Phys. J. C},
	year = 2020,
	volume = 80,
	number = 12,
	pages = "1213",
}

@article{Zhang:2010iea,
    author = "Zhang, Jin-Mei and Wang, Guo-Li",
    title = "{Strong Decays of the Radial Excited States $B(2S)$ and $D(2S)$}",
    eprint = "1001.1773",
    archivePrefix = "arXiv",
    primaryClass = "hep-ph",
    doi = "10.1016/j.physletb.2010.01.016",
    journal = "Phys. Lett. B",
    volume = "684",
    pages = "221--223",
    year = "2010"
}

@article{LHCb:2013jjb,
    author = "Aaij, R and others",
    collaboration = "LHCb",
    title = "{Study of $D_J$ meson decays to $D^+\pi^-$, $D^0 \pi^+$ and $D^{*+}\pi^-$ final states in pp collision}",
    eprint = "1307.4556",
    archivePrefix = "arXiv",
    primaryClass = "hep-ex",
    reportNumber = "CERN-PH-EP-2013-122, LHCB-PAPER-2013-026",
    doi = "10.1007/JHEP09(2013)145",
    journal = "JHEP",
    volume = "09",
    pages = "145",
    year = "2013"
}

@article{BaBar:2010zpy,
    author = "del Amo Sanchez, P. and others",
    collaboration = "BaBar",
    title = "{Observation of new resonances decaying to $D\pi$ and $D^*\pi$ in inclusive $e^+e^-$ collisions near $\sqrt{s}=$10.58 GeV}",
    eprint = "1009.2076",
    archivePrefix = "arXiv",
    primaryClass = "hep-ex",
    reportNumber = "BABAR-PUB-10-020, SLAC-PUB-14246",
    doi = "10.1103/PhysRevD.82.111101",
    journal = "Phys. Rev. D",
    volume = "82",
    pages = "111101",
    year = "2010"
}

@article{Na:2015kha,
    author = "Na, Heechang and Bouchard, Chris M. and Lepage, G. Peter and Monahan, Chris and Shigemitsu, Junko",
    collaboration = "HPQCD",
    title = "{$B \rightarrow D l \nu$ form factors at nonzero recoil and extraction of $|V_{cb}|$}",
    eprint = "1505.03925",
    archivePrefix = "arXiv",
    primaryClass = "hep-lat",
    doi = "10.1103/PhysRevD.93.119906",
    journal = "Phys. Rev. D",
    volume = "92",
    number = "5",
    pages = "054510",
    year = "2015",
    note = "[Erratum: Phys.Rev.D 93, 119906 (2016)]"
}

@article{Yao:2019vty,
    author = "Yao, De-Liang and Fernandez-Soler, Pedro and Guo, Feng-Kun and Nieves, Juan",
    title = "{New parametrization of the form factors in $\bar{B}\to D\ell\bar{\nu}_\ell$ decays}",
    eprint = "1906.00727",
    archivePrefix = "arXiv",
    primaryClass = "hep-ph",
    doi = "10.1103/PhysRevD.101.034014",
    journal = "Phys. Rev. D",
    volume = "101",
    number = "3",
    pages = "034014",
    year = "2020"
}

@article{Isgur:1990yhj,
    author = "Isgur, Nathan and Wise, Mark B.",
    title = "{Weak transition form factors between heavy mesons}",
    reportNumber = "UTPT-90-01, CALT-68-1608",
    doi = "10.1016/0370-2693(90)91219-2",
    journal = "Phys. Lett. B",
    volume = "237",
    pages = "527--530",
    year = "1990"
}

@article{Polchinski:2002jw,
    author = "Polchinski, Joseph and Strassler, Matthew J.",
    title = "{Deep inelastic scattering and gauge / string duality}",
    eprint = "hep-th/0209211",
    archivePrefix = "arXiv",
    reportNumber = "NSF-ITP-02-139, UW-PT-02-23",
    doi = "10.1088/1126-6708/2003/05/012",
    journal = "JHEP",
    volume = "05",
    pages = "012",
    year = "2003"
}

@article{Erlich:2005qh,
    author = "Erlich, Joshua and Katz, Emanuel and Son, Dam T. and Stephanov, Mikhail A.",
    title = "{QCD and a holographic model of hadrons}",
    eprint = "hep-ph/0501128",
    archivePrefix = "arXiv",
    reportNumber = "SLAC-PUB-10965, WM-05-101, INT-PUB-05-02",
    doi = "10.1103/PhysRevLett.95.261602",
    journal = "Phys. Rev. Lett.",
    volume = "95",
    pages = "261602",
    year = "2005"
}

@article{FolcoCapossoli:2019imm,
    author = "Folco Capossoli, Eduardo and Mart{\'\i}n Contreras, Miguel Angel and Li, Danning and Vega, Alfredo and Boschi-Filho, Henrique",
    title = "{Hadronic spectra from deformed AdS backgrounds}",
    eprint = "1903.06269",
    archivePrefix = "arXiv",
    primaryClass = "hep-ph",
    doi = "10.1088/1674-1137/44/6/064104",
    journal = "Chin. Phys. C",
    volume = "44",
    number = "6",
    pages = "064104",
    year = "2020"
}

@article{Abidin:2009hr,
    author = "Abidin, Zainul and Carlson, Carl E.",
    title = "{Nucleon electromagnetic and gravitational form factors from holography}",
    eprint = "0903.4818",
    archivePrefix = "arXiv",
    primaryClass = "hep-ph",
    doi = "10.1103/PhysRevD.79.115003",
    journal = "Phys. Rev. D",
    volume = "79",
    pages = "115003",
    year = "2009"
}

@article{Aharony:1999ti,
    author = "Aharony, Ofer and Gubser, Steven S. and Maldacena, Juan Martin and Ooguri, Hirosi and Oz, Yaron",
    title = "{Large N field theories, string theory and gravity}",
    eprint = "hep-th/9905111",
    archivePrefix = "arXiv",
    reportNumber = "CERN-TH-99-122, HUTP-99-A027, LBNL-43113, RU-99-18, UCB-PTH-99-16, LBL-43113",
    doi = "10.1016/S0370-1573(99)00083-6",
    journal = "Phys. Rept.",
    volume = "323",
    pages = "183--386",
    year = "2000"
}

@article{Grigoryan:2007vg,
    author = "Grigoryan, Hovhannes R. and Radyushkin, Anatoly V.",
    title = "{Form Factors and Wave Functions of Vector Mesons in Holographic QCD}",
    eprint = "hep-ph/0703069",
    archivePrefix = "arXiv",
    reportNumber = "JLAB-THY-07-624",
    doi = "10.1016/j.physletb.2007.05.044",
    journal = "Phys. Lett. B",
    volume = "650",
    pages = "421--427",
    year = "2007"
}

@article{Boschi-Filho:2002xih,
    author = "Boschi-Filho, Henrique and Braga, Nelson R. F.",
    title = "{Gauge / string duality and scalar glueball mass ratios}",
    eprint = "hep-th/0212207",
    archivePrefix = "arXiv",
    doi = "10.1088/1126-6708/2003/05/009",
    journal = "JHEP",
    volume = "05",
    pages = "009",
    year = "2003"
}

@article{Brodsky:2007hb,
    author = "Brodsky, Stanley J. and de Teramond, Guy F.",
    title = "{Light-Front Dynamics and AdS/QCD Correspondence: The Pion Form Factor in the Space- and Time-Like Regions}",
    eprint = "0707.3859",
    archivePrefix = "arXiv",
    primaryClass = "hep-ph",
    reportNumber = "SLAC-PUB-12554",
    doi = "10.1103/PhysRevD.77.056007",
    journal = "Phys. Rev. D",
    volume = "77",
    pages = "056007",
    year = "2008"
}

@article{Colangelo:1998fi,
    author = "Colangelo, P. and De Fazio, F. and Paver, N.",
    editor = "Narison, Stephan",
    title = "{Universal Isgur-Wise form-factors from QCD sum rules in HQET}",
    eprint = "hep-ph/9810478",
    archivePrefix = "arXiv",
    reportNumber = "BARI-TH-98-311",
    doi = "10.1016/S0920-5632(99)00167-X",
    journal = "Nucl. Phys. B Proc. Suppl.",
    volume = "74",
    pages = "222--226",
    year = "1999"
}

@article{Bernard:1993ct,
    author = "Bernard, Claude W. and Shen, Yue and Soni, Amarjit",
    title = "{Isgur-Wise function on the lattice}",
    eprint = "hep-lat/9312038",
    archivePrefix = "arXiv",
    reportNumber = "BUHEP-93-27",
    doi = "10.1016/0920-5632(94)90424-3",
    journal = "Nucl. Phys. B Proc. Suppl.",
    volume = "34",
    pages = "483--485",
    year = "1994"
}

@article{Monahan:2017uby,
    author = "Monahan, Christopher J and Na, Heechang and Bouchard, Chris M and Lepage, G Peter and Shigemitsu, Junko",
    title = "{$B_s \to D_s \ell \nu$ Form Factors and the Fragmentation Fraction Ratio $f_s/f_d$}",
    eprint = "1703.09728",
    archivePrefix = "arXiv",
    primaryClass = "hep-lat",
    doi = "10.1103/PhysRevD.95.114506",
    journal = "Phys. Rev. D",
    volume = "95",
    number = "11",
    pages = "114506",
    year = "2017"
}

@inproceedings{Chen:2024hln,
    author = "Chen, Chen",
    collaboration = "LHCb",
    title = "{$b \to c l \bar{\nu}$ decays at LHCb}",
    booktitle = "{58th Rencontres de Moriond on QCD and High Energy Interactions}",
    eprint = "2405.08953",
    archivePrefix = "arXiv",
    primaryClass = "hep-ex",
    month = "5",
    year = "2024"
}

@article{Hao:2022ibj,
    author = "Hao, Wei and Lu, Yu and Wang, En",
    title = "{The assignments of the $B_s$ mesons within the screened potential model and $^3P_0$ model}",
    eprint = "2212.10068",
    archivePrefix = "arXiv",
    primaryClass = "hep-ph",
    doi = "10.1140/epjc/s10052-023-11689-3",
    journal = "Eur. Phys. J. C",
    volume = "83",
    number = "6",
    pages = "520",
    year = "2023"
}

@article{Colangelo:2008us,
    author = "Colangelo, P. and De Fazio, F. and Giannuzzi, Floriana and Jugeau, F. and Nicotri, S.",
    title = "{Light scalar mesons in the soft-wall model of AdS/QCD}",
    eprint = "0807.1054",
    archivePrefix = "arXiv",
    primaryClass = "hep-ph",
    reportNumber = "BARI-TH-08-593",
    doi = "10.1103/PhysRevD.78.055009",
    journal = "Phys. Rev. D",
    volume = "78",
    pages = "055009",
    year = "2008"
}

@article{Bowler:2002zh,
    author = "Bowler, K. C. and Douglas, G. and Kenway, R. D. and Lacagnina, G. N. and Maynard, C. M.",
    collaboration = "UKQCD",
    title = "{Semileptonic decays of heavy mesons and the Isgur-Wise function in quenched lattice QCD}",
    eprint = "hep-lat/0202029",
    archivePrefix = "arXiv",
    reportNumber = "EDINBURGH-2002-01",
    doi = "10.1016/S0550-3213(02)00462-5",
    journal = "Nucl. Phys. B",
    volume = "637",
    pages = "293--310",
    year = "2002"
}

@article{FermilabLatt:2015ilb,
    author = "Bailey, Jon A. and others",
    collaboration = "Fermilab Lattice, MILC",
    title = "{B{\textrightarrow}D{\ensuremath{\ell}}{\ensuremath{\nu}} form factors at nonzero recoil and |V$_{cb}$| from 2+1-flavor lattice QCD}",
    eprint = "1503.07237",
    archivePrefix = "arXiv",
    primaryClass = "hep-lat",
    reportNumber = "FERMILAB-PUB-15-107-T",
    doi = "10.1103/PhysRevD.92.034506",
    journal = "Phys. Rev. D",
    volume = "92",
    number = "3",
    pages = "034506",
    year = "2015"
}

@article{Luke:1990eg,
    author = "Luke, Michael E.",
    title = "{Effects of subleading operators in the heavy quark effective theory}",
    reportNumber = "HUTP-90-A051",
    doi = "10.1016/0370-2693(90)90568-Q",
    journal = "Phys. Lett. B",
    volume = "252",
    pages = "447--455",
    year = "1990"
}

@article{LHCb:2020gnv,
    author = "Aaij, Roel and others",
    collaboration = "LHCb",
    title = "{Observation of a New Excited $D^+_s$ Meson in $B^0 \rightarrow D^- D^+ K^+ \pi^-$ Decays}",
    eprint = "2011.09112",
    archivePrefix = "arXiv",
    primaryClass = "hep-ex",
    reportNumber = "CERN-EP-2020-203, LHCb-PAPER-2020-034",
    doi = "10.1103/PhysRevLett.126.122002",
    journal = "Phys. Rev. Lett.",
    volume = "126",
    number = "12",
    pages = "122002",
    year = "2021"
}

@article{Sun:2009tg,
    author = "Sun, Zhi-Feng and Liu, Xiang",
    title = "{Newly observed D(sJ)(3040) and the radial excitations of P-wave charmed-strange mesons}",
    eprint = "0909.1658",
    archivePrefix = "arXiv",
    primaryClass = "hep-ph",
    doi = "10.1103/PhysRevD.80.074037",
    journal = "Phys. Rev. D",
    volume = "80",
    pages = "074037",
    year = "2009"
}

@article{BaBar:2009rro,
    author = "Aubert, Bernard and others",
    collaboration = "BaBar",
    title = "{Study of D(sJ) decays to D*K in inclusive e+ e- interactions}",
    eprint = "0908.0806",
    archivePrefix = "arXiv",
    primaryClass = "hep-ex",
    reportNumber = "SLAC-PUB-13747, BABAR-PUB-09-024",
    doi = "10.1103/PhysRevD.80.092003",
    journal = "Phys. Rev. D",
    volume = "80",
    pages = "092003",
    year = "2009"
}

@article{Bordone:2025jur,
    author = "Bordone, Marzia and Gubernari, Nico and Jung, Martin and van Dyk, Danny",
    title = "{Challenging $ {\overline{B}}_{(s)}\to {D}_{(s)}^{\left(\ast \right)} $ form factors with the heavy quark expansion}",
    eprint = "2507.03569",
    archivePrefix = "arXiv",
    primaryClass = "hep-ph",
    reportNumber = "CERN-TH-2025-092, EOS-2025-03, IPPP/25/25, P3H-25-032, SI-HEP-2025-10, ZU-TH 35/25",
    doi = "10.1007/JHEP11(2025)051",
    journal = "JHEP",
    volume = "11",
    pages = "051",
    year = "2025"
}

@article{Yadav:2024zrn,
    author = "Yadav, Ajay Kumar and Sahoo, Suchismita",
    title = "{Binwise exploration of vector couplings in Bs{\textrightarrow}Ds(*){\ensuremath{\tau}}{\ensuremath{\nu}}{\textasciimacron}{\ensuremath{\tau}} decays}",
    eprint = "2410.06100",
    archivePrefix = "arXiv",
    primaryClass = "hep-ph",
    doi = "10.1016/j.nuclphysb.2025.117164",
    journal = "Nucl. Phys. B",
    volume = "1020",
    pages = "117164",
    year = "2025"
}

@article{Ahmed:2023pod,
    author = "Ahmed, Hiwa A. and Chen, Yidian and Huang, Mei",
    title = "{D(s)- mesons semileptonic form factors in four-flavor holographic QCD}",
    eprint = "2309.06156",
    archivePrefix = "arXiv",
    primaryClass = "hep-ph",
    doi = "10.1103/PhysRevD.109.026008",
    journal = "Phys. Rev. D",
    volume = "109",
    number = "2",
    pages = "026008",
    year = "2024"
}

@article{Momeni:2022gqb,
    author = "Momeni, S. and Saghebfar, M.",
    title = "{Semileptonic $D$ meson decays to the vector, axial vector and scalar mesons in Hard-Wall AdS/QCD correspondence}",
    doi = "10.1140/epjc/s10052-022-10413-x",
    journal = "Eur. Phys. J. C",
    volume = "82",
    number = "5",
    pages = "473",
    year = "2022"
}

@article{Ahmady:2013cga,
    author = "Ahmady, M. and Campbell, R. and Lord, S. and Sandapen, R.",
    title = "{Predicting the $B -> \rho$ form factors using AdS/QCD Distribution Amplitudes for the $\rho$ meson}",
    eprint = "1308.3694",
    archivePrefix = "arXiv",
    primaryClass = "hep-ph",
    doi = "10.1103/PhysRevD.88.074031",
    journal = "Phys. Rev. D",
    volume = "88",
    number = "7",
    pages = "074031",
    year = "2013"
}

@article{Chow:1996xf,
    author = "Chow, Chi-Keung",
    title = "{Large N(c) universality of the baryon Isgur-Wise form-factor: The Group theoretical approach}",
    eprint = "hep-ph/9601248",
    archivePrefix = "arXiv",
    reportNumber = "CLNS-96-1392",
    doi = "10.1103/PhysRevD.54.873",
    journal = "Phys. Rev. D",
    volume = "54",
    pages = "873--876",
    year = "1996"
}

@article{Lee:2002ka,
    author = "Lee, Jong-Phil and Park, Gye T.",
    title = "{Subleading Isgur-Wise function of Lambda(b) to Lambda(c1) using QCD sum rules}",
    eprint = "hep-ph/0210097",
    archivePrefix = "arXiv",
    doi = "10.1016/S0370-2693(02)03161-1",
    journal = "Phys. Lett. B",
    volume = "552",
    pages = "185--197",
    year = "2003"
}

@article{Sadzikowski:1993iv,
    author = "Sadzikowski, M. and Zalewski, K.",
    title = "{Isgur-Wise functions from the MIT bag model}",
    eprint = "hep-ph/9304319",
    archivePrefix = "arXiv",
    reportNumber = "TPJU-9-93",
    doi = "10.1007/BF01562560",
    journal = "Z. Phys. C",
    volume = "59",
    pages = "677--681",
    year = "1993"
}

@article{Brodsky:2006uqa,
    author = "Brodsky, Stanley J. and de Teramond, Guy F.",
    title = "{Hadronic spectra and light-front wavefunctions in holographic QCD}",
    eprint = "hep-ph/0602252",
    archivePrefix = "arXiv",
    reportNumber = "SLAC-PUB-11716",
    doi = "10.1103/PhysRevLett.96.201601",
    journal = "Phys. Rev. Lett.",
    volume = "96",
    pages = "201601",
    year = "2006"
}

@article{Brodsky:2008pg,
    author = "Brodsky, Stanley J. and de Teramond, Guy F.",
    editor = "Zichichi, Antonino",
    title = "{AdS/CFT and Light-Front QCD}",
    eprint = "0802.0514",
    archivePrefix = "arXiv",
    primaryClass = "hep-ph",
    reportNumber = "SLAC-BPUB-13107",
    doi = "10.1142/9789814293242_0008",
    journal = "Subnucl. Ser.",
    volume = "45",
    pages = "139--183",
    year = "2009"
}

@article{Brodsky:1973kr,
    author = "Brodsky, Stanley J. and Farrar, Glennys R.",
    title = "{Scaling Laws at Large Transverse Momentum}",
    reportNumber = "SLAC-PUB-1290",
    doi = "10.1103/PhysRevLett.31.1153",
    journal = "Phys. Rev. Lett.",
    volume = "31",
    pages = "1153--1156",
    year = "1973"
}

@article{Witten:1998qj,
    author = "Witten, Edward",
    title = "{Anti-de Sitter space and holography}",
    eprint = "hep-th/9802150",
    archivePrefix = "arXiv",
    primaryClass = "hep-th",
    doi = "10.4310/ATMP.1998.v2.n2.a2",
    journal = "Adv. Theor. Math. Phys.",
    volume = "2",
    pages = "253--291",
    year = "1998"
}

@book{Grozin:2004,
    author = "Grozin, Andrey G.",
    title = "{Heavy Quark Effective Theory}",
    doi = "10.1007/b97728",
    isbn = "978-3-540-20692-7",
    publisher = "Springer",
    series = "Springer Tracts in Modern Physics",
    volume = "201",
    year = "2004"
}

@article{Klebanov:1999tb,
    author = "Klebanov, Igor R. and Witten, Edward",
    title = "{AdS / CFT correspondence and symmetry breaking}",
    eprint = "hep-th/9905104",
    archivePrefix = "arXiv",
    primaryClass = "hep-th",
    doi = "10.1016/S0550-3213(99)00453-8",
    journal = "Nucl. Phys. B",
    volume = "556",
    pages = "89--114",
    year = "1999"
}
\end{document}